\documentclass[twocolumn,switch]{article} 

\usepackage{preprint}

\usepackage{amssymb}
\usepackage{amsmath}
\usepackage{latexsym}
\usepackage[ruled,vlined]{algorithm2e} 
\usepackage{array}
\usepackage{tabularx}
\usepackage{booktabs} 
\usepackage{graphicx}

\graphicspath{{./}{../}}
 
\usepackage{url}
\usepackage{xcolor}
\definecolor{newcolor}{rgb}{.8,.349,.1}

\usepackage[capitalize]{cleveref}
\crefname{section}{sec.}{secs.}
\Crefname{section}{Sec.}{Secs.}
\crefname{table}{tab.}{tabs.}
\Crefname{table}{Tab.}{Tabs.}
\crefname{figure}{fig.}{figs.}
\Crefname{figure}{Fig.}{Figs.}
\crefname{equation}{eq.}{eqs.}
\Crefname{equation}{Eq.}{Eqs.}
\crefname{algorithm}{alg.}{algs.}
\Crefname{algorithm}{Alg.}{Algs.}

\usepackage{natbib}
\title{Domain Agnostic Image-to-image Translation using Low-Resolution Conditioning}

\usepackage{authblk}
\author[1]{Mohamed Abid}
\author[1]{Arman Afrasiyabi}
\author[1]{Ihsen Hedhli}
\author[1]{Jean-Fran\c{c}ois Lalonde}
\author[1,2\thanks{\tt{christian.gagne@gel.ulaval.ca}}]{Christian Gagn\'e}

\affil[1]{Institute Intelligence and Data (IID), Universit\'e Laval}
\affil[2]{Canada CIFAR AI Chair, Mila -- Quebec AI Institute}

\begin{document}                                                         
\onecolumn

\maketitle

\begin{abstract} 
Generally, image-to-image translation (i2i) methods aim at learning mappings across domains with the assumption that the images used for translation share content (e.g., pose) but have their own domain-specific information (a.k.a.\ style). Conditioned on a target image, such methods extract the target style and combine it with the source image content, keeping coherence between the domains. In our proposal, we depart from this traditional view and instead consider the scenario where the target domain is represented by a very low-resolution (LR) image, proposing a domain-agnostic i2i method for fine-grained problems, where the domains are related. More specifically, our domain-agnostic approach aims at generating an image that combines visual features from the source image with low-frequency information (e.g. pose, color) of the LR target image. To do so, we present a novel approach that relies on training the generative model to produce images that both share distinctive information of the associated source image and correctly match the LR target image when downscaled. We validate our method on the CelebA-HQ and AFHQ datasets by demonstrating improvements in terms of visual quality. Qualitative and quantitative results show that when dealing with intra-domain image translation, our method generates realistic samples compared to state-of-the-art methods such as StarGAN v2. Ablation studies also reveal that our method is robust to changes in color, it can be applied to out-of-distribution images, and it allows for manual control over the final results. 
\end{abstract}

\vfill

Under consideration in \emph{Computer Vision and Image Understanding}.

\newpage
\twocolumn

%!TEX root = main.tex
\section{Introduction}

Image-to-image translation (i2i) methods seek to learn a mapping between two related domains, thereby ``translating'' an image from one domain to another while preserving some information from the original. These methods have been used for various applications in computer vision, such as colorization \citep{color}, super-resolution \citep{srgan, Wang_2018_ECCV_Workshops}, medical imaging \citep{armanious2020medgan}, and photorealistic image synthesis \citep{zhang2017stackgan}, achieving promising results both in terms of visual quality and diversity. Here, one key idea is to enforce the model to preserve the content of the image while modifying its style. This inspired many works such as \cite{huang2018multimodal} and \cite{DRIT} to disentangle the feature space into 1) a domain-specific space for the style; and 2) a shared space for the content---this allowed more diverse multimodal generation. 

Recent methods such as StarGAN \citep{stargan,starganv} unified the process in a single framework that works across many domains. These methods also introduced the problem of \emph{conditional} image-to-image translation which aims at conditioning this process using a specific image from the target domain. As defined by \cite{CItoI}, conditional i2i ``\emph{requires that the generated image should inherit some domain-specific features of the conditional image from the target domain}''. Traditionally, methods aim to extract the style of a target image and merge it with the content of a source image to generate an image that shares information from both the source and target. Additionally, domain-specific information also guides the learning process. For instance, to translate an image from one domain to another domain, StarGAN \citep{stargan,starganv} requires domain supervision in the form of a domain label.

\begin{figure} 
    \centering
    \footnotesize
    \setlength{\tabcolsep}{1pt}
    \begin{tabular}{ccc} 
    \includegraphics[width=0.50\linewidth, angle=0]{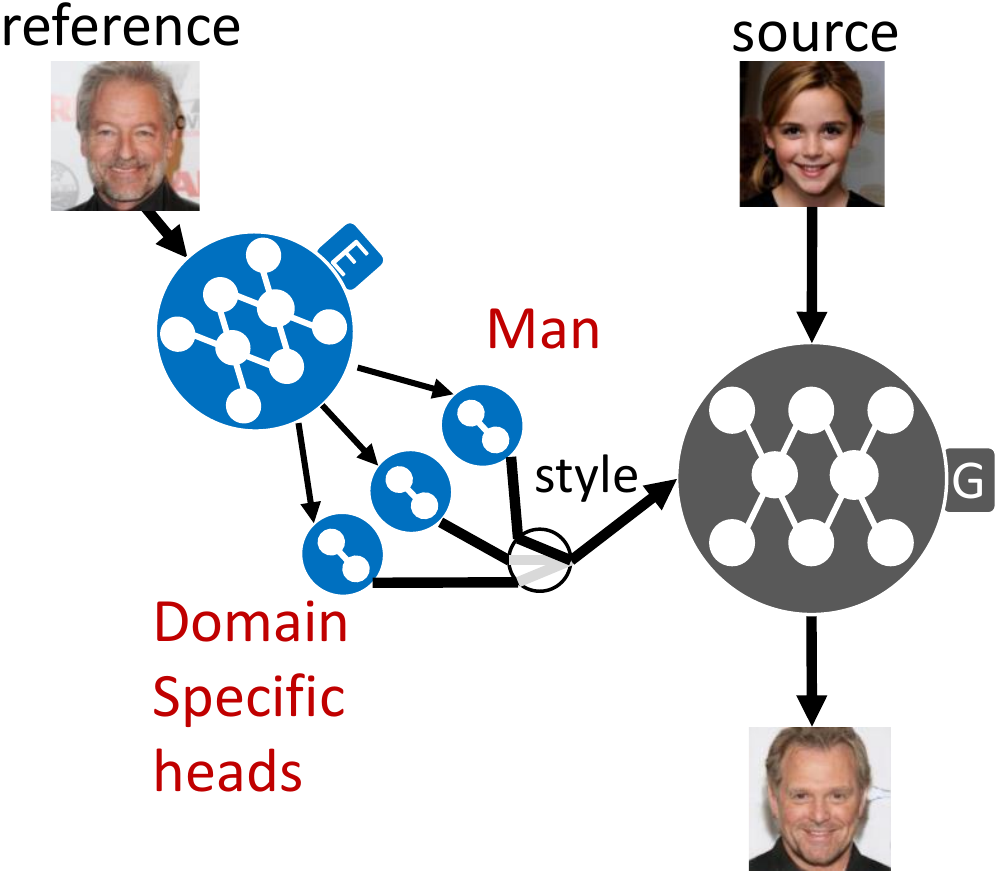} & 
    \includegraphics[width=0.50\linewidth, angle=0]{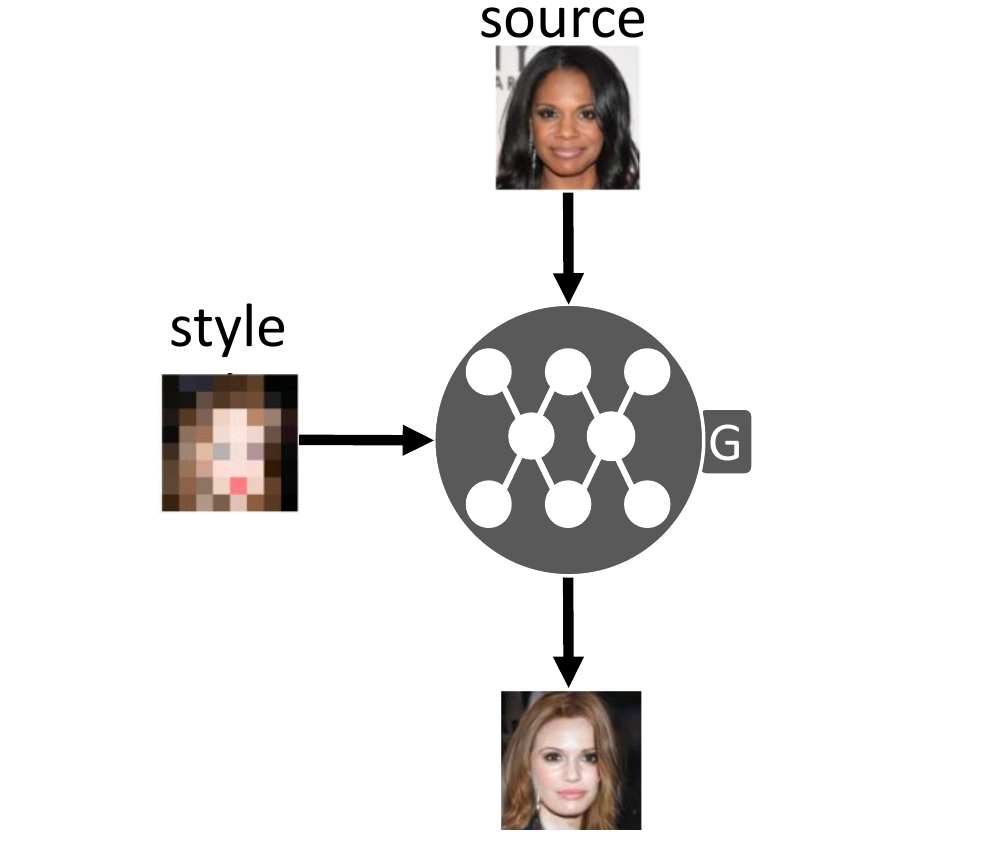} &   \\ 
    (a) Domain-based StarGAN & 
    (b) Domain agnostic (ours)    \\       
    \end{tabular}
    \vspace{0.5em}
    \caption{Comparison of StarGAN \citep{stargan,starganv} as a conventional method with our domain-agnostic approach. While domain-based StarGAN (a) aims at extracting domain-specific style using a style encoder $E$ from the reference image, our domain-agnostic approach (b) uses the low-resolution conditioning directly without requiring knowledge of a domain label. }
    \label{fig:star_ours}
\end{figure}
  
\begin{figure} 
\begin{center}
%\fbox{\rule{0pt}{2in} \rule{.9\linewidth}{0pt}}
\includegraphics[width=\linewidth]{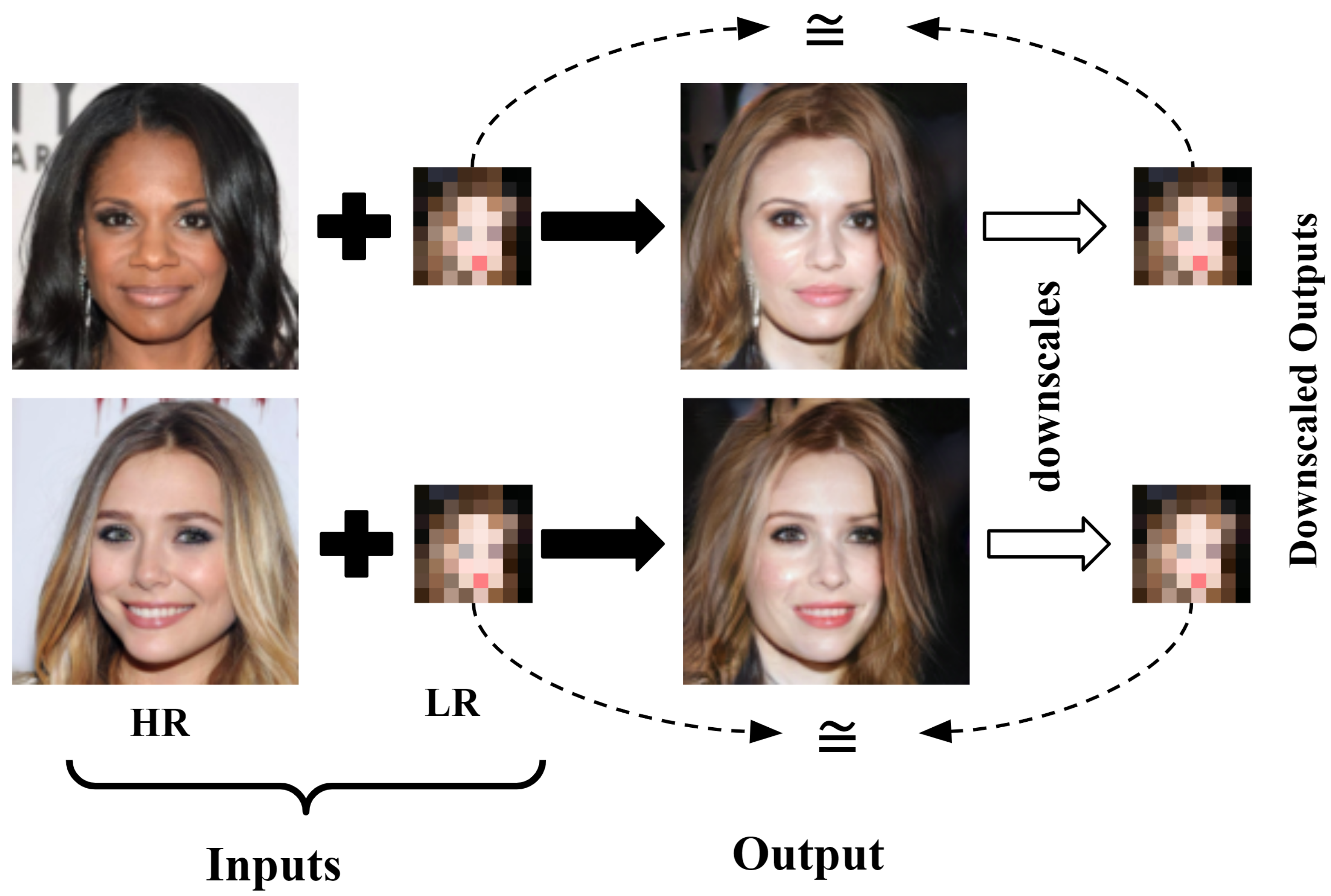}
\end{center}
  \caption{Illustration of the proposed approach on CelebA-HQ dataset: our method learned to map the inputs high-resolution image (HR, source) and low-resolution (LR, target) image to generate a HR output that preserves the identity of the source image and stays faithful to the structure of the LR target.}
\label{fig:first}
\end{figure}

In this work, we depart from this traditional view of conditional i2i and propose a novel perspective for domain-agnostic i2i. 
% for fine-grained scenarios, where the domains are associated. 
In particular, we consider the scenario where the ``content'' of an image is defined by its low-frequency information, while the ``style'' is defined by the high-frequency. 
% Indeed, low-frequencies in images represent pixel values that are changing slowly over space (content in our case), while high-frequency content represents pixel values that are rapidly changing in space (details in our case).  
% To preserve only the low-frequency information from a given image we make use of a very low-resolution (LR) version of that image, therefore making it impossible to extract any ``style'' information. 
Specifically, from an input source and a downscaled version of the target, we wish to recover an image that contains high-frequency information (details) from the source image while downscaling to the low-resolution (LR) target image. 
% Coming back to the above definition, the ``domain-specific features'' is the low-frequency information (e.g. pose, color distribution) contained in the LR of the target. 
Contrarily to previous work, no target domain label is provided to the model: domain information is obtained directly from the target image itself. As illustrated in \cref{fig:star_ours}-(a), conventional cross-domain supervised methods such as StarGAN \citep{stargan,starganv} use a style encoder network $E$ to extract a style vector from the reference image, which is then injected in the generator network $G$. 
In comparison and as shown in \cref{fig:star_ours}-(b), our domain-agnostic approach is a fully unsupervised network across the domains, and our proposed algorithm only employs a single generator $G$ network. As illustrated in \cref{fig:first}, the resulting learned mapping can generate an image that shares distinctive features from both the high-resolution (HR) source image and the LR target image in an unsupervised manner.

The main contribution of this paper is to define a novel domain-agnostic framework for fine-grained problems that can learn a mapping between images while being conditioned on the low-frequency information represented by a very low-resolution version of the target. Particularly, our method can perform i2i translation between domains without injecting label information, in a fully unsupervised manner. 
% Besides, our learning algorithm does not employ an additional style encoding network, hence it contains fewer parameters compared to the transitional methods \todo{Are results validating this statement given somewhere? I don't remember have seen it, if we don't validate this strictly, we should remove the sentence.} 
We demonstrate that our approach can effectively be used to transfer useful features (e.g., pose, spatial structure) from downscaled targets where details are completely blurred to the point of being visually unrecognizable. 
When evaluating our approach on the CelebA-HQ and AFHQ datasets, we show that our framework results in realistic samples according to the FID and LPIPS scores on both datasets. We also provide more evidence with our ablation study on a different aspect of our model. These extensive experiments demonstrate that our method can generate results that are photorealistic and that convincingly fuse information from both the HR source and LR target images. Finally, performed ablations show how our approach results in better image synthesis compared to StarGAN \citep{stargan,starganv} under the same training settings.

\section{Related work}
  
Generative adversarial networks (GANs) \citep{goodfellow} are prominent generative methods with demonstrated results in various applications in computer vision, including image generation \citep{Biggan,karras2019style,stylegan2}, super-resolution \citep{srgan,Wang_2018_ECCV_Workshops,kim2022region} and image-to-image translation \citep{hu2022style, huang2018multimodal, kim2022instaformer, isola2017image,DRIT,msgan,cycleGan, kim2022instaformer}. 

Some work has striven to improve sample quality and diversity, notably through theoretical breakthroughs in terms of defining loss functions that provide more stable training \citep{Wgan,wgangp,lsgan,EnegryGan,hoyer2022hrda} and encourage diversity in the generated images \citep{Greg,diverseGan}. Architectural innovations also played a crucial role in these advancements \citep{kim2019u, Biggan,starganv,ProGanCelebA,liu2017unsupervised,yang2022unsupervised}. For instance, \cite{liu2017unsupervised} makes use of an attention layer that allows it to focus on long-range dependencies present in the image. Spectral normalization \citep{miyato2018spectral} stabilizes the network, which also translates into having high-quality samples. Since our work sits at the intersection of reference-guided image synthesis and super-resolution, the remainder of this section focuses on these two branches exclusively.

%Also, recent methods \cite{StyleGan,starganv} introduced a mapping network designed to take a normalized latent vector and output a disentangled latent vector with same size which fed to into intermediate adaptive instance normalization \cite{Adain} layers of the generator borrowed from the style transfer literature. 

\paragraph{Reference-guided image synthesis} 
Also dubbed ``conditional'' \citep{CItoI}, reference-guided i2i translation methods seek to learn a mapping from a source to a target domain while being conditioned on a specific image instance belonging to the target domain. In this case, some methods attempt to preserve the ``content'' of the source image (e.g., identity, pose) and apply the ``style'' (e.g., hair/skin color) of the target. Inspired by the mapping network of StyleGAN \citep{karras2019style}, recent methods \citep{stargan,starganv,huang2018multimodal} make use of a style encoder to extract the style of a target image and feed it to the generator, typically via adaptive instance normalization (AdaIN) \citep{huang2017arbitrary}. All these methods have the built-in assumption that the resolution of the source and target images are the same. In this work, the ``style'' of the target is preserved by conditioning on a LR version of the same target image. 

% Although these methods are conditioned solely on target images from the designated domain with the same resolution, the proposed technique is conditioned on a very low resolution version of the target image.

\paragraph{Super-resolution}
Our approach bears some resemblance to super-resolution methods, which are aiming to learn a mapping from LR to a plausible HR images \citep{srgan,dong2014learning}, trained over ground truth HR images. More closely related are the so-called reference-guided super-resolution methods, which, in addition to the LR input image, also accept an additional HR images for guiding the generation process \citep{CrossCSR,srntt}. The critical difference is that, in this paper, we do not aim to recover the ground truth HR image. Rather, our method extracts distinctive information from the source image (HR) and generates an image which simultaneously: 1) contains this information; and 2) downscales to the target LR image.

\section{Method}
\begin{figure} 
    \centering
    \footnotesize
    \setlength{\tabcolsep}{1pt}
    \begin{tabular}{ccc} 
    \includegraphics[width=0.45\linewidth, angle=0]{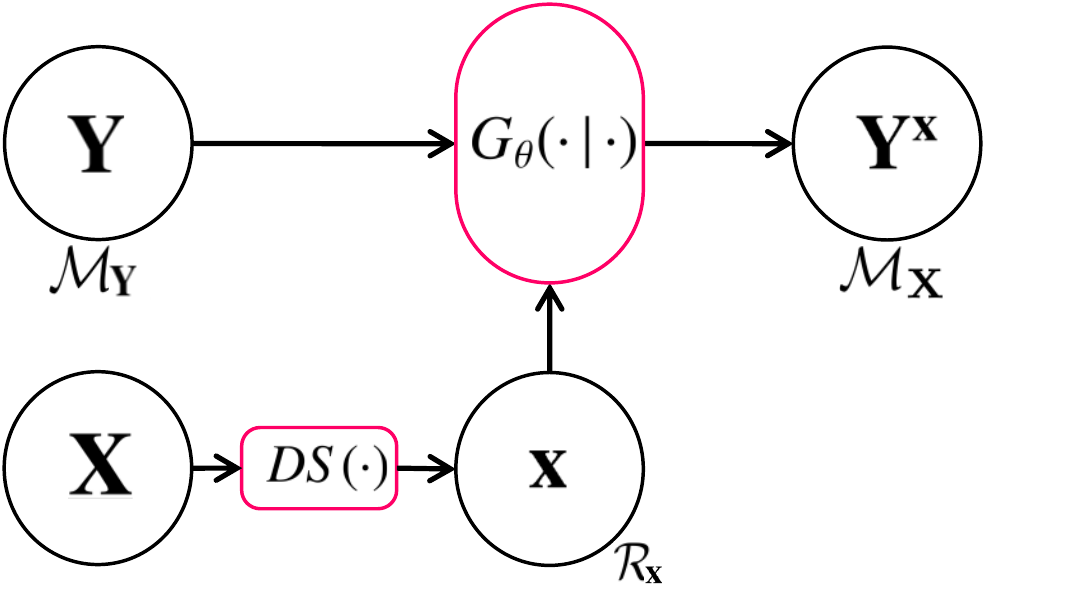} & 
    \includegraphics[width=0.45\linewidth, angle=0]{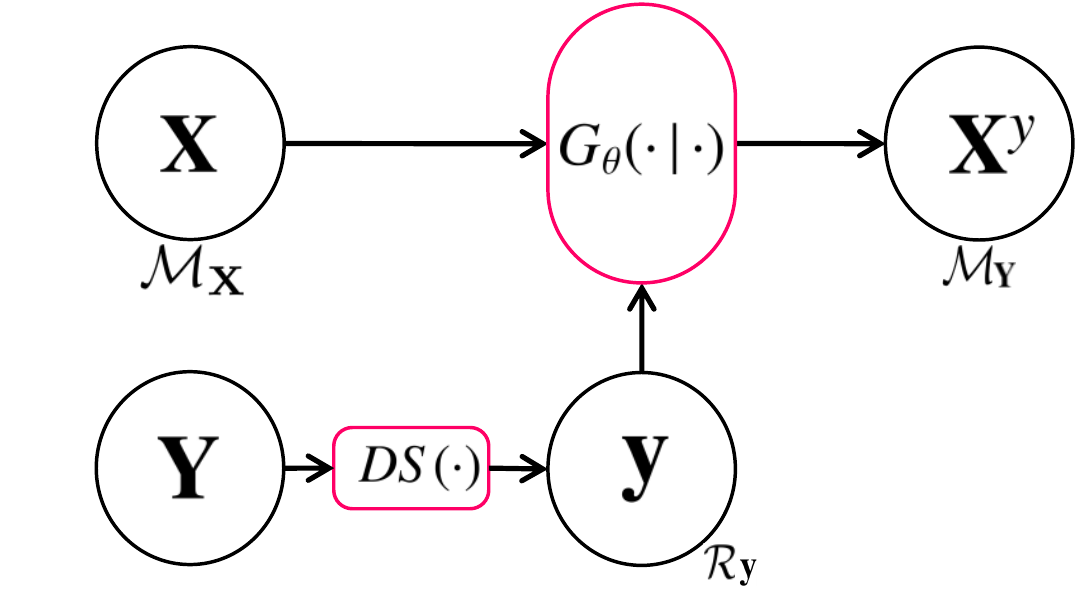} &    
    \\ \ \\
    % (a) [Translation of HR source image $\mathbf{Y}$ from $\mathcal{M}_\mathbf{Y}$ to $\mathcal{M}_\mathbf{X}$ guided by LR target $\mathbf{x}$. & 
    % (b) Translation of HR source image $\mathbf{X}$ from $\mathcal{M}_\mathbf{X}$ to $\mathcal{M}_\mathbf{Y}$ guided by LR target $\mathbf{y}$. \\  
    (a) $\mathbf{X}$ in the style of  $\mathbf{Y}$ & 
    (b) $\mathbf{Y}$ in the style of  $\mathbf{X}$ \\  
    \end{tabular}
    % \vspace{0.5em}
    \caption{Illustration of our domain-agnostic method. Generator $ G_{\theta}$ is trained to achieve a high-resolution mapping of a $\mathbf{Y}$ (resp. $\mathbf{X}$) from its current domain $\mathcal{M}_\mathbf{Y}$ (resp. $\mathcal{M}_\mathbf{X}$) to the target domain $\mathcal{M}_\mathbf{X}$ (resp. $\mathcal{M}_\mathbf{Y}$), conditioned on $\mathbf{x}$, a low-resolution version of $\mathbf{X}$ (resp. $\mathbf{y}$ of $\mathbf{Y}$). 
    (a) Translation of high-resolution image $\mathbf{Y}$ from $\mathcal{M}_\mathbf{Y}$ to $\mathcal{M}_\mathbf{X}$ guided by $\mathbf{x}$, a low-resolution version of the target.
    (b) Generation of $\mathbf{X}^\mathbf{y}$, that correspond to $\mathbf{X}$ in the style of $\mathbf{Y}$.
    % \todo{(CG) I find this figure to be unclear and inconsistent, it should be revised.}
    % b) Translation of HR image $\mathbf{X}$ from $\mathcal{M}_\mathbf{X}$ to $\mathcal{M}_\mathbf{Y}$ guided by LR target $\mathbf{y}$.
    }% 
    \label{fig:space_illustration}
    % \label{fig:idea_illustration}
\end{figure}

To define the target domain, we are given a low-resolution (LR) version of the target image $\mathbf{x} \in \mathbb{R}^{m \times n}$, and we define the associated $\mathcal{R}_\mathbf{x}$ subspace of LR images as:
\begin{equation}
\mathcal{R}_\mathbf{x}=\left\{\forall i, \mathbf{x}_i \in \mathbb{R}^{m \times n}:\left\|\mathbf{x}_i - \mathbf{x}\right\|_{p} \leq \epsilon\right\}  \,.
\label{Rx}
\end{equation}
We consider an $\epsilon$ close to zero, so that each subspace contains only the LR images $\mathbf{x}_i$ that are highly similar  to the LR target $\mathbf{x}$ according to a given norm $p$ (here $p=1$).
We are also defining a target subspace $\mathcal{M}_\mathbf{X}$ in the source image manifold that includes all images that are included in $\mathcal{R}_\mathbf{x}$ when downscaled as LR images:
\begin{equation}
\mathcal{M}_\mathbf{X} = \left\{\forall j, \mathbf{X}_j \in \mathbb{R}^{M \times N}: \mathrm{DS}(\mathbf{X}_j) \in \mathcal{R}_\mathbf{x} \right\}   \,,
\end{equation}
where $\mathbf{X}_j$ is an image in the high-resolution (HR) space, and $\mathrm{DS}(\cdot)$ is a downscaling operator. As in \cite{berthelot2020creating}, average pooling is used for downscaling with $\mathrm{DS}(\cdot)$. Therefore, for each subspace $\mathcal{M}_\mathbf{X}$, there exists a corresponding subspace $\mathcal{R}_\mathbf{x}$ in the LR space related by  $\mathrm{DS}(\cdot)$. 
As illustrated in \cref{fig:space_illustration}, given two pairs $(\mathbf{X} \in \mathcal{M}_\mathbf{X},\,\mathbf{x}=\mathrm{DS}(\mathbf{X}))$ and $(\mathbf{Y} \in \mathcal{M}_\mathbf{Y},\,\mathbf{y} =\mathrm{DS}(\mathbf{Y}))$, our goal is to learn a domain agnostic function $G$ that translates the source images from one subspace $\mathcal{M}_\mathbf{X}$ to another subspace $\mathcal{M}_\mathbf{Y}$. These two subspaces are defined according to information provided by their LR target version counterpart (i.e., $\mathcal{R}_\mathbf{x}$ and $\mathcal{R}_\mathbf{y}$, respectively):
\begin{equation} 
    \begin{split}
    G:& \, \mathcal{M}_\mathbf{Y} \times \mathcal{R}_\mathbf{x} \mapsto \mathcal{M}_\mathbf{X}, \;  G_{\theta}(\mathbf{Y}\,|\,\mathbf{x})\,, \\
    G:& \, \mathcal{M}_\mathbf{X} \times \mathcal{R}_\mathbf{y} \mapsto \mathcal{M}_\mathbf{Y}, \;  G_{\theta}(\mathbf{X}\,|\,\mathbf{y})
    \,.
    \end{split}  
\end{equation}

% \begin{figure} 
%     \centering
%     \footnotesize
%     \setlength{\tabcolsep}{1pt}
%     \begin{tabular}{ccc} 
%     \includegraphics[width=0.35\linewidth, angle=0]{fig/fig_ab_a.pdf} & 
%     \includegraphics[width=0.35\linewidth, angle=0]{fig/fig_ab_b.pdf} &   \\ 
%     (a) $\mathbf{X}$ in the style of  $\mathbf{Y}$ & 
%     (b) $\mathbf{Y}$ in the style of  $\mathbf{X}$ \\  
%     \end{tabular}
%     % \vspace{0.5em}
%     \caption{Conceptual illustration of our proposed training algorithm using parameterized generator $G_{\theta}$ and non-parametric function downsample function $DS(\cdot)$: (a) synthesising $\mathbf{X}^\mathbf{y}$ which is $\mathbf{X}$ in the style of  $\mathbf{Y}$, (b) synthesising $\mathbf{Y}^{\mathbf{x}}$ that is $\mathbf{Y}$ in the style of  $\mathbf{x}$. During the training procedure of $G_{\theta}$, we tune the parameters using both (a) and (b).
%     }
%     \label{fig:idea_illustration}
% \end{figure}

As shown in \cref{fig:space_illustration} (b), in practice, we follow a straightforward learning algorithm for our proposed domain-agnostic approach. For example, our domain-agnostic approach outputs $\mathbf{X}^{y}$ which is the translation of HR input image $\mathbf{X}$ in the style of $\mathbf{y}$. To this end, we employ the generator $ G_{\theta}$ network parameterized by $\theta$.

Following conventional GAN terminology \citep{goodfellow2020generative}, parameterized function $G_{\theta}$ is a generator (simplified as $G$ hereinafter) which aims at translating a source image from a subspace to a target subspace while preserving distinctive source features. Training $G$ relies on a discriminator $D$ that plays two roles: 1) to classify whether the generated images are fake or real, pushing $G$ to generate samples on the natural image manifold; and 2) to evaluate whether the translated source image is part of the right target domain according to the downsampling constraint.

\begin{figure}[t]
\centering 
\includegraphics[width=.9\linewidth]{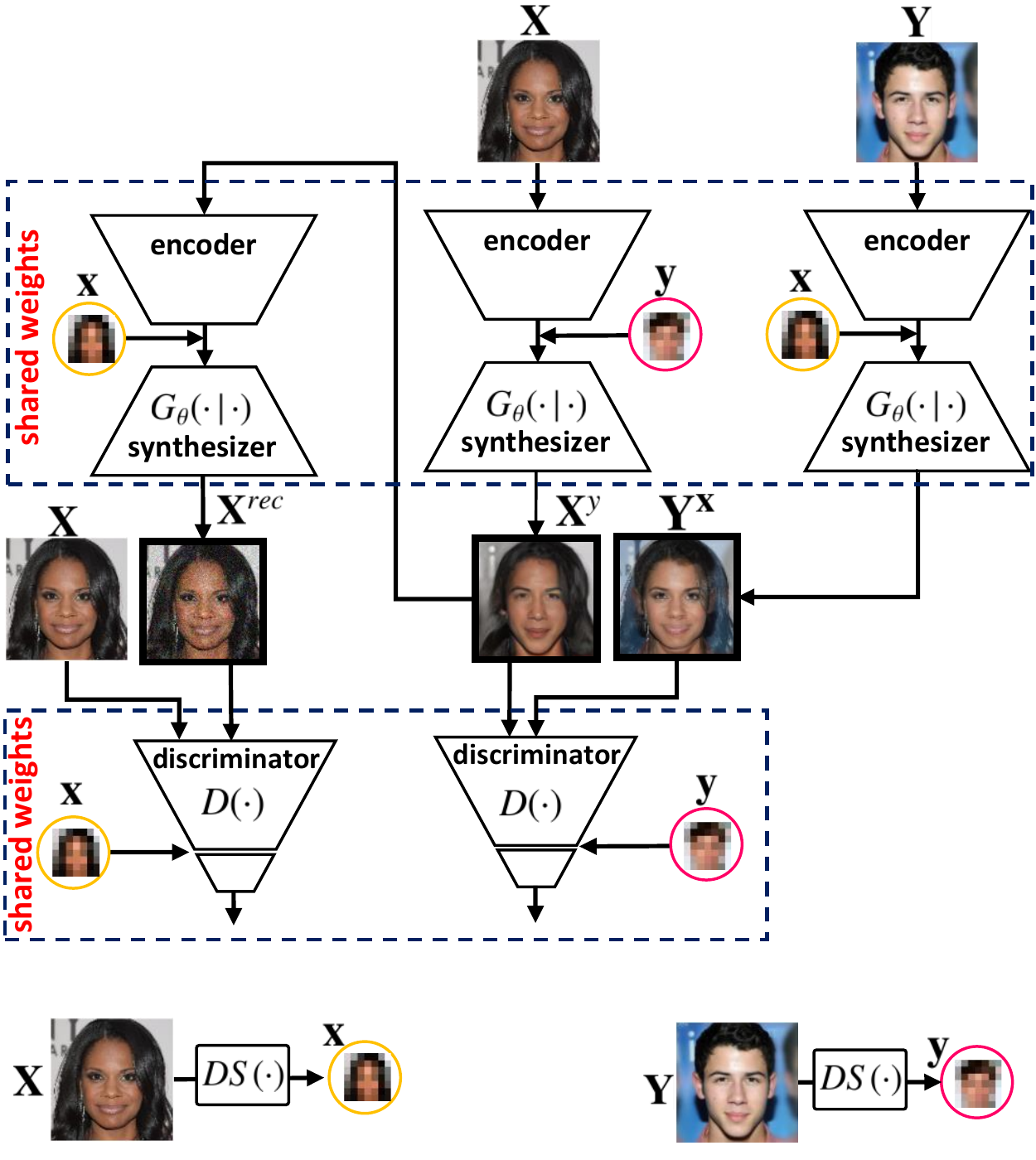} 
\caption{Schematic overview of our training algorithm to obtain $\mathbf{X}^{\mathbf{y}}$ (resp. $\mathbf{Y}^{\mathbf{x}}$) which corresponds to translating image $\mathbf{X}$ (resp. $\mathbf{Y}$) to the target domain of image $\mathbf{Y}$ (resp. $\mathbf{X}$) based on its low resolution version $\mathbf{y}$ (resp. $\mathbf{x}$). Generator $G_{\theta}$ translates high resolution input images using low resolution versions of the target domain images. Besides, network $D_{\phi}$ aims at discriminating between $\mathbf{X}^{\mathbf{y}}$ and $\mathbf{Y}^{\mathbf{x}}$, and between the reconstructed $\mathbf{X}^{rec} = G_{\theta}(\mathbf{X}^{\mathbf{y}}\,|\,\mathbf{x})$ version of $\mathbf{X}^{\mathbf{y}}$ and the original $\mathbf{X}$. The networks inside each dashed box (labeled with shared weights) are the same. The low resolution versions of the input images $\mathbf{x}$ and $\mathbf{y}$ are obtained using a non-parametric function $DS(\cdot)$ function applied on $\mathbf{X}$ and $\mathbf{Y}$, respectively. 
%
% First, the generator $ G_{\theta}$ synthesis the fake image $\mathbf{X}^{fake}$ conditioned on the low resolution of $\mathbf{y} = DS(\mathbf{Y})$. Here, $DS(\cdot)$ is the down-sample function. Next, we extract fake version $\mathbf{Y}^{fake}$ of image $\mathbf{Y}$  conditioned on $\mathbf{x} = DS(\mathbf{X})$, and the discriminator aims at distinguishing between $\mathbf{X}^{fake}$ and $\mathbf{Y}^{fake}$. At the same time, we train the discriminator to classify between $\mathbf{X}$ and the reconstructed version of $\mathbf{X}^{fake}$.
%
}
\label{fig:architecture_}
\end{figure} 
%!TEX root = main.tex 
\begin{algorithm}[t]
    \SetAlgoLined
    {\small
        \KwData{Generator $G_{\theta}(\cdot\,|\,\cdot)$, discriminator $D_{\phi}(\cdot)$, downsampling function $DS(\cdot)$, training dataset $\mathcal{X}_\text{\emph{train}}$, learning rate $\eta$, 
         maximum number of training iterations $t^\text{\emph{max}}$.  
        }
        % \KwResult{... 
        % }  
        \For{$t=1,\ldots,t^\text{\emph{max}}$} {
        {\color{teal}\# \footnotesize Pick a random pair of samples and their low-resolution} \ \\ 
         sample $\mathbf{X}\in\mathcal{X}_\text{\emph{train}}$ and $\mathbf{Y}\in\mathcal{X}_\text{\emph{train}}$\\  
        %  {\color{teal} \footnotesize \# compute the LRs}  \\
         $\mathbf{x} = DS(\mathbf{X})\ ; \quad \mathbf{y} = DS(\mathbf{Y})$\\ 
        {\color{teal} \footnotesize \# Feedforward samples through generator} \\
         $\mathbf{X}^{\mathbf{y}} = G_{\theta}\big(\mathbf{X}\,|\,\mathbf{y}\big)\ ; \quad \ \ \ \mathbf{Y}^{\mathbf{x}} = G_{\theta}\big(\mathbf{Y}\,|\,\mathbf{x}\big)$\\
        $\mathbf{X}^{rec} = G_{\theta}\big(\mathbf{X}^{\mathbf{y}}\,|\,\mathbf{x}\big)\ ; \quad  \mathbf{Y}^{rec} = G_{\theta}\big(\mathbf{Y}^{\mathbf{x}}\,|\,\mathbf{y}\big)$\\
        Compute adversarial $\mathcal{L}_\mathrm{adv}$ loss in the form of Eq.~\ref{overall_adversarial_loss} \\
        % D_{\phi}(real_samps, real_low, fake_samps)
        % $\mathcal{L}_\mathrm{adv} = \mathcal{L}_\mathrm{adv}(\mathbf{X},\mathbf{Y}^\mathbf{x}) + \mathcal{L}_\mathrm{adv}(\mathbf{X},\mathbf{X}^{rec}) + \mathcal{L}_\mathrm{adv}(\mathbf{Y},\mathbf{X}^\mathbf{y}) + \mathcal{L}_\mathrm{adv}(\mathbf{Y},\mathbf{Y}^{rec})\,.$\\
        % \todo{(CG) I am unsure that computing the adversarial loss on $\mathcal{L}_\mathrm{adv}(\mathbf{X},\mathbf{X}^{rec})$ and $\mathcal{L}_\mathrm{adv}(\mathbf{Y},\mathbf{Y}^{rec})$ is appropriate, we need to check whether it is really required (if not, better remove it).}\\
        
        % {\color{teal} \footnotesize \# compute the reconstruction losses } \\    
        {\color{teal} \footnotesize \# Update discriminator with adversarial loss} \\
        $\phi \xleftarrow{} \phi - \eta \nabla_{\phi}\mathcal{L}_\mathrm{adv}$\\
        Compute cycle loss $\mathcal{L}_\mathrm{cyc}$ using Eq.~\ref{L_cyc} \\
        % $\mathcal{L}_\mathrm{cyc} = \mathcal{L}_\mathrm{cyc}(\mathbf{X},\mathbf{Y}^{\mathbf{x}}) + \mathcal{L}_\mathrm{cyc}(\mathbf{X},\mathbf{X}^{rec}) + \mathcal{L}_\mathrm{cyc}(\mathbf{Y}, \mathbf{Y}^{\mathbf{x}}) + \mathcal{L}_\mathrm{cyc}(\mathbf{Y},\mathbf{Y}^{rec})$\,.\\
        % }
        {\color{teal} \footnotesize \# Compute reconstruction loss using $l_1$ norm} \\
        $\mathcal{L}_\mathrm{rec} = \|\mathbf{X}-\mathbf{X}^{rec}\|_1 + \|\mathbf{Y}-\mathbf{Y}^{rec}\|_1$ \\  
        % \todo{(CG) There is something odd here with the $\mathcal{L}_\mathrm{cyc}(\mathbf{X},\mathbf{X}^{rec})$ and $\mathcal{L}_\mathrm{cyc}(\mathbf{Y},\mathbf{Y}^{rec})$ elements of the cycle loss are doing much the same thing than the reconstruction loss.}\\
        {\color{teal} \footnotesize \# Update generator with cycle and reconstruction losses} \\
        $\theta \xleftarrow{} \theta - \eta \nabla_{\theta} (\mathcal{L}_\mathrm{cyc} + \mathcal{L}_\mathrm{rec})$ 
        } 
    }  
    \caption{Domain-agnostic image-to-image translation from low-resolution conditionning}
    \label{algo:training}
\end{algorithm}

\begin{figure} %[t] 
\centering
%\fbox{\rule{0pt}{2in} \rule{.9\linewidth}{0pt}}
\includegraphics[width=\linewidth]{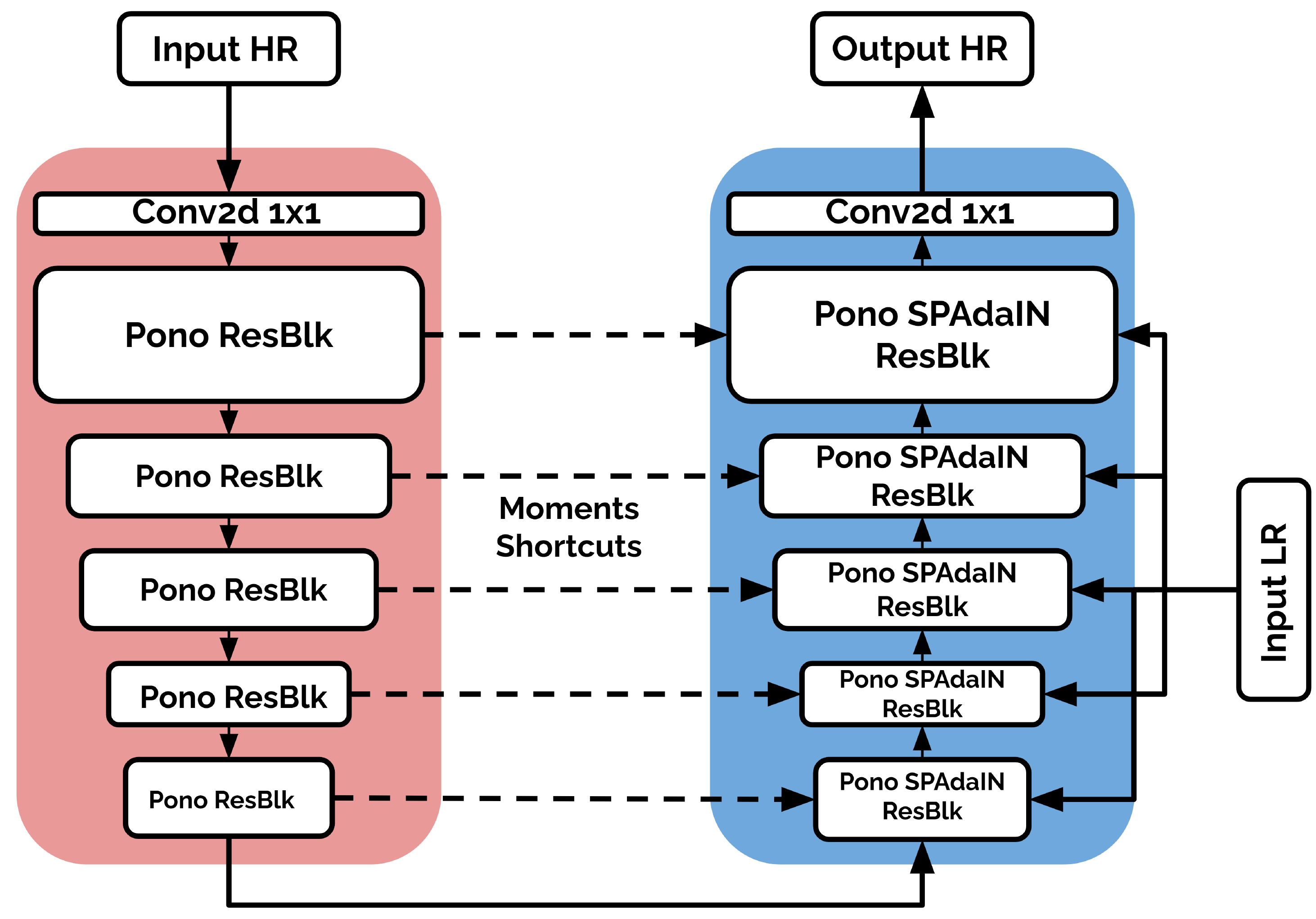} 
\caption{Generator architecture: the encoding part (in red) is used for extracting structural information in the source image, which is merged in the decoder (in blue) with the incoming features from the decoding output and the low resolution targets.}
\label{fig:generator}
\end{figure}

\subsection{Training Objectives}
% \subsection{Architecture}
% \begin{figure*}[ht]
%     \centering
%     \subfloat[Pono residual blocks]{\includegraphics[width=0.42 \textwidth]{fig/ponoresblock2.pdf}
%     \label{fig:res a}}%
%     \qquad
%     \subfloat[Pono SPAdaIn residual blocks]{\includegraphics[width=0.42\textwidth]{fig/ponoSpadainResblock.pdf}
%     \label{fig:res b}}%
%     \caption{Residual blocks used in the generator $G$: (a) Encoding pono residual blocks, used to extract and compress the features coming from the source image and pass it to the decoding part of the generator through moments shortcuts and as compressed features  ; and (b) Decoding pono SPAdaIn residual blocks takes as input extracted features from the encoding part, and also upsample the LR target image and pass it to the SPAdaIn layer \citep{NeuralPose} }%
%     \label{fig:resblocks}%
% \end{figure*}
For achieving these two discrimination roles, the GAN is trained according to the following minimax optimization:
\begin{equation}
\min_{G} \max_{D} ~ \mathcal{L}_\mathrm{adv} + \lambda_\mathrm{cyc}\,\mathcal{L}_\mathrm{cyc}\,,
\label{eq:fullobj}
\end{equation}
where $\mathcal{L}_\mathrm{adv}$ is an adversarial loss used to ensure that we are generating plausible natural images, $\mathcal{L}_\mathrm{cyc}$ is the cycle consistency loss ensuring that the translated images are kept on the correct subspace while carrying the right features, and $\lambda_\mathrm{cyc}$ is a hyperparameter to achieve the right balance between these two losses. The process of translating $\mathbf{X}$ in the style of $\mathbf{Y}$ is illustrated in \cref{fig:architecture_}. First, we translate $\mathbf{X}$ (resp. $\mathbf{Y}$) using the parameterized generator, that is $\mathbf{X}^\mathbf{y} =  G_{\theta}(\mathbf{X}\,|\,\mathbf{y})$  (resp. $\mathbf{Y}^\mathbf{x} =  G_{\theta}(\mathbf{Y}\,|\,\mathbf{x})$). Then, we compute the reconstructed $\mathbf{X}^{rec} = G_{\theta}(\mathbf{X}^{y}\,|\,\mathbf{x})$ version of $\mathbf{X}$. Finally, the resulting images are feedforwarded twice through discriminator $D_{\phi}(\cdot)$: 1) to determine which image between $\mathbf{X}^{\mathbf{y}}$ and $\mathbf{Y}^{\mathbf{x}}$ are conditionned on the low resolution $\mathbf{x}$, and 2) to discriminate between $\mathbf{X}$ and its corresponding reconstruction $\mathbf{X}^{rec}$. 
% Note that \cref{fig:architecture_} only presents three forward pass (for clarity) through generator to compute $\mathbf{X}^\mathbf{y}$, and our learning algorithm also performs the fourth forward to compute $\mathbf{Y}^\mathbf{x}$.    

At each training iteration, four forward passes are done with the generator: the first two for generating $\mathbf{X}^\mathbf{y}=G_{\theta}(\mathbf{X}\,|\,\mathbf{y})$ and $\mathbf{Y}^\mathbf{x} =G_{\theta}(\mathbf{Y}\,|\,\mathbf{x})$, while the other two are for cycle consistency (i.e., $G_{\theta}(G_{\theta}(\mathbf{X}\,|\,\mathbf{y})\,|\,\mathbf{x}) \approx \mathbf{X}$ and $G_{\theta}( G_{\theta}(\mathbf{Y}\,|\,\mathbf{x})\,|\,\mathbf{y})\approx\mathbf{Y}$). The discriminator is used to make sure that the generated samples are from the designated subspace.

\paragraph{Adversarial loss} 
Following \cite{berthelot2020creating}, we provide the discriminator with the absolute difference $\mathbf{d}_{\mathbf{X}\rightarrow \mathcal{M}_\mathbf{Y}}\in\mathbb{R}^{m\times n}$ between the downscaled version of the generated image $\mathrm{DS}( G_{\theta}(\mathbf{X}\,|\,\mathbf{y}))$ and the LR target $\mathbf{y}=\mathrm{DS}(\mathbf{Y})$:
\begin{equation}
    \mathbf{d}_{\mathbf{X}\rightarrow \mathcal{M}_\mathbf{Y}} = \frac{\left|\mathbf{y}-\left\lfloor r\, \mathrm{DS}( G_{\theta}(\mathbf{X}\,|\,\mathbf{y}))\right\rfloor\right|}{r}\,,
    \label{eq:ste}
\end{equation}
where $r$ is the color resolution. As in \cite{berthelot2020creating}, we round the downscaled image to its nearest color resolution ($r=2/255$, since pixel values are in $[-1,1]$) to avoid unstable optimization caused by exceedingly large weights to measure small pixel value differences. A straight-through estimator \citep{steEstimator} is employed to pass the gradient through the rounding operation in eq.~\ref{eq:ste}. The discriminator, therefore, takes as inputs:
\begin{equation}
    \left\{
    \begin{array}{ll}
    D_{\phi}(\mathbf{Y},\mathbf{0}) & \text{for real samples,} \\
    D_{\phi}\left( G_{\theta}(\mathbf{X}\,|\,\mathbf{y}),\,\mathbf{d}_{\mathbf{X}\rightarrow \mathcal{M}_\mathbf{Y}} \right) & \text {otherwise.}
    \end{array} \right.
    \label{eq:disInput}
\end{equation}
Here, $\mathbf{0}$ is an all-zeros $m\times n$ image difference, since the downscaled version of $\mathbf{Y}$ is exactly $\mathbf{y}$. However, for fake samples, the absolute difference $\mathbf{d}_{\mathbf{X}\rightarrow \mathcal{M}_\mathbf{Y}}$ depends on how close is the generator to the designated subspace, $\mathcal{M}_\mathbf{Y}$ in our example. Both networks $G$ and $D$ are trained via the resulting adversarial loss:
\begin{align}\label{adversarial_loss}
  \mathcal{L}_\mathrm{adv}(\mathbf{X},\mathbf{Y}) = & \ \log D_{\phi}(\mathbf{X},\mathbf{0}) + \log D_{\phi}(\mathbf{Y},\mathbf{0})\nonumber\\
  & + \log \left(1-D_{\phi}( G_{\theta}(\mathbf{X}\,|\,\mathrm{DS}(\mathbf{Y})),\,\mathbf{d}_{\mathbf{X}\rightarrow \mathcal{M}_\mathbf{Y}} )\right)\,.
\end{align}
In the same manner, we compute the adversarial loss between $(\mathbf{X},\mathbf{Y}^\mathbf{x})$, $(\mathbf{X},\mathbf{X}^{rec})$, $(\mathbf{Y},\mathbf{X}^\mathbf{y})$, and $(\mathbf{Y},\mathbf{Y}^{rec})$ pairs, respectively. Finally, we compute the overall adversarial loss $\mathcal{L}_\mathrm{adv}$ in the following form:  
\begin{align}\label{overall_adversarial_loss} \nonumber
\mathcal{L}_\mathrm{adv} = \mathcal{L}_\mathrm{adv}(\mathbf{X},\mathbf{Y}^\mathbf{x}) + \mathcal{L}_\mathrm{adv}(\mathbf{X},\mathbf{X}^{rec})  \ \\ + \ \mathcal{L}_\mathrm{adv}(\mathbf{Y},\mathbf{X}^\mathbf{y}) + \mathcal{L}_\mathrm{adv}(\mathbf{Y},\mathbf{Y}^{rec})\,.
\end{align}

\paragraph{Cycle consistency loss}  To make sure that generator $G$ preserves the distinctive features available in the source image, we employ the cycle consistency constraint \citep{isola2017image,liu2017unsupervised,cycleGan} in both directions, each time by changing the LR target to specify the designated subspace: 
\begin{align} \label{L_cyc}
\mathcal{L}_\mathrm{cyc}(\mathbf{X},\mathbf{Y}) = & \ \|\mathbf{X}- G_{\theta}( G_{\theta}(\mathbf{X}\,|\,\mathrm{DS}(\mathbf{Y}))\,|\,\mathrm{DS}(\mathbf{X}))\|_1\nonumber\\
    & + \|\mathbf{Y}- G_{\theta}( G_{\theta}(\mathbf{Y}\,|\,\mathrm{DS}(\mathbf{X})\,|\,\mathrm{DS}(\mathbf{Y}))\|_1\,.
\end{align}
Likewise, we compute the cycle loss between $(\mathbf{X},\mathbf{Y}^{\mathbf{x}})$, $(\mathbf{X},\mathbf{X}^{rec})$, $(\mathbf{X},\mathbf{X}^{rec})$, $(\mathbf{Y}, \mathbf{Y}^{\mathbf{x}})$, and $(\mathbf{Y},\mathbf{Y}^{rec})$. Finally, we compute the overall adversarial loss $\mathcal{L}_\mathrm{cyc}$ in the following form: 
\begin{align}\label{overall_cycle_loss} \nonumber
\mathcal{L}_\mathrm{cyc} = \mathcal{L}_\mathrm{cyc}(\mathbf{X},\mathbf{Y}^{\mathbf{x}})   + \mathcal{L}_\mathrm{cyc}(\mathbf{X},\mathbf{X}^{rec}) \ \\ + \mathcal{L}_\mathrm{cyc}(\mathbf{Y}, \mathbf{Y}^{\mathbf{x}}) + \mathcal{L}_\mathrm{cyc}(\mathbf{Y},\mathbf{Y}^{rec})
\end{align}

This cycle consistency loss encourages the generator to identify the shared and invariant information between every two subspaces for preserving it during translation.
\Cref{algo:training} illustrates how we feedforward through the geneator $G_{\theta}$ and the discriminator $D_{\phi}$ and compute the losses during our proposed domain-agnostic i2i approach.

\subsection{Architecture}
Most current image-to-image translation models \citep{stargan, huang2018multimodal} rely on Adaptive Instance Normalization (AdaIN) \citep{huang2017arbitrary, karras2019style} to transfer the style from a reference image to a source image. However, in our work the hypothesis of content and style is not suitable since the LR image contains information on both style (e.g., colors) and content (e.g., pose). Thus, our generator adapts source image to the content and style of the LR image through the use of spatially adaptive instance normalization (SPAdaIN) \citep{NeuralPose}.
% Therefore, the use of AdaIN~\citep{Adain} in our generator does not server our need, since, it s shown in \citep{NeuralPose} that AdaIN, is not suitable for shapes and pose transfer, due to the lack of learnable parameters, which will only push the network to adopt the present shapes and pose of the input high resolution image.

% Since our task, requires that the generated images when down-scaled is spatially aligned with the corresponding LR of the targeted subspaces. 

\begin{figure}[t]
    \centering
    \footnotesize
    \includegraphics[width=0.8\linewidth, angle=0]{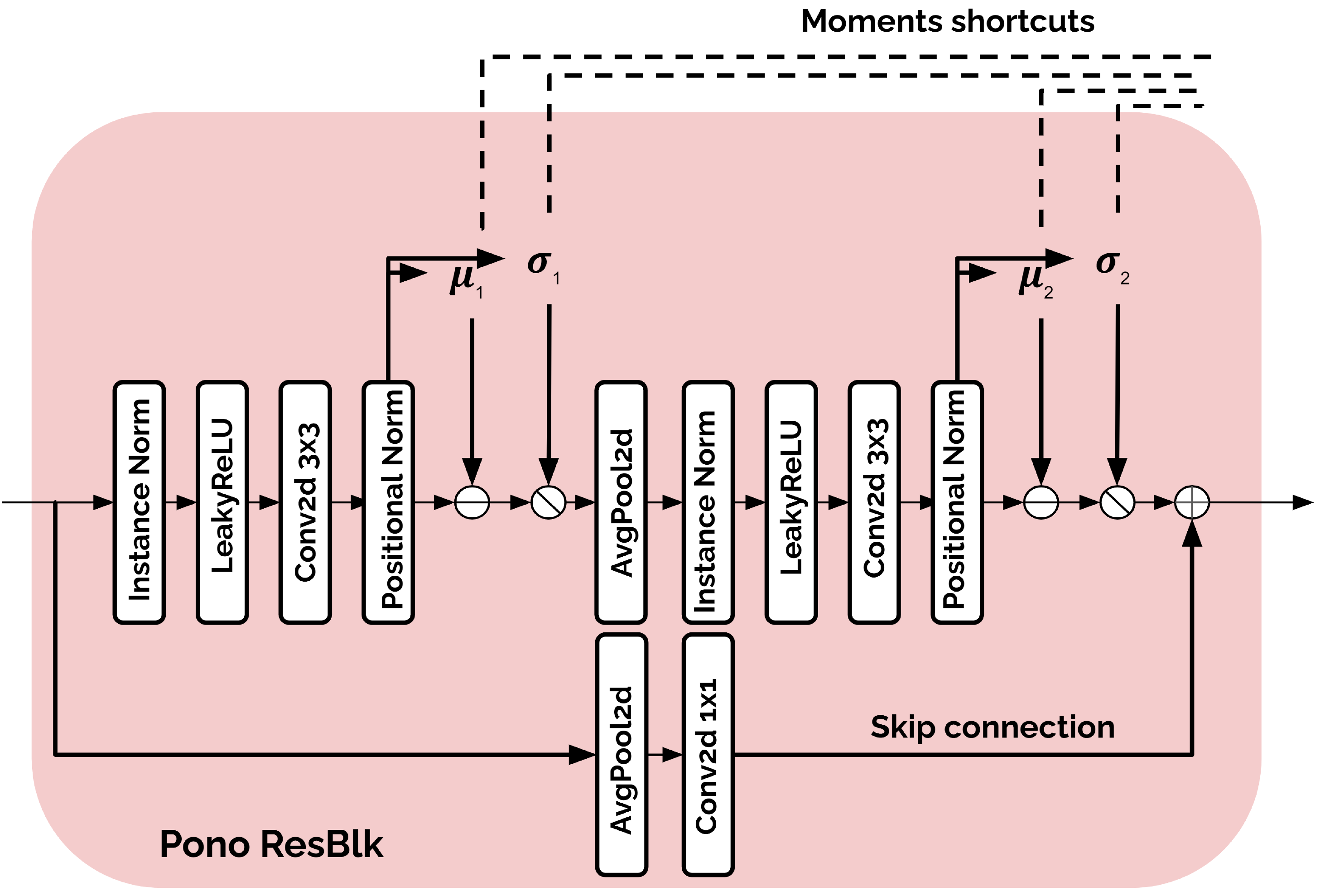}\\
    (a) Pono residual blocks\\
    \includegraphics[width=0.8\linewidth, angle=0]{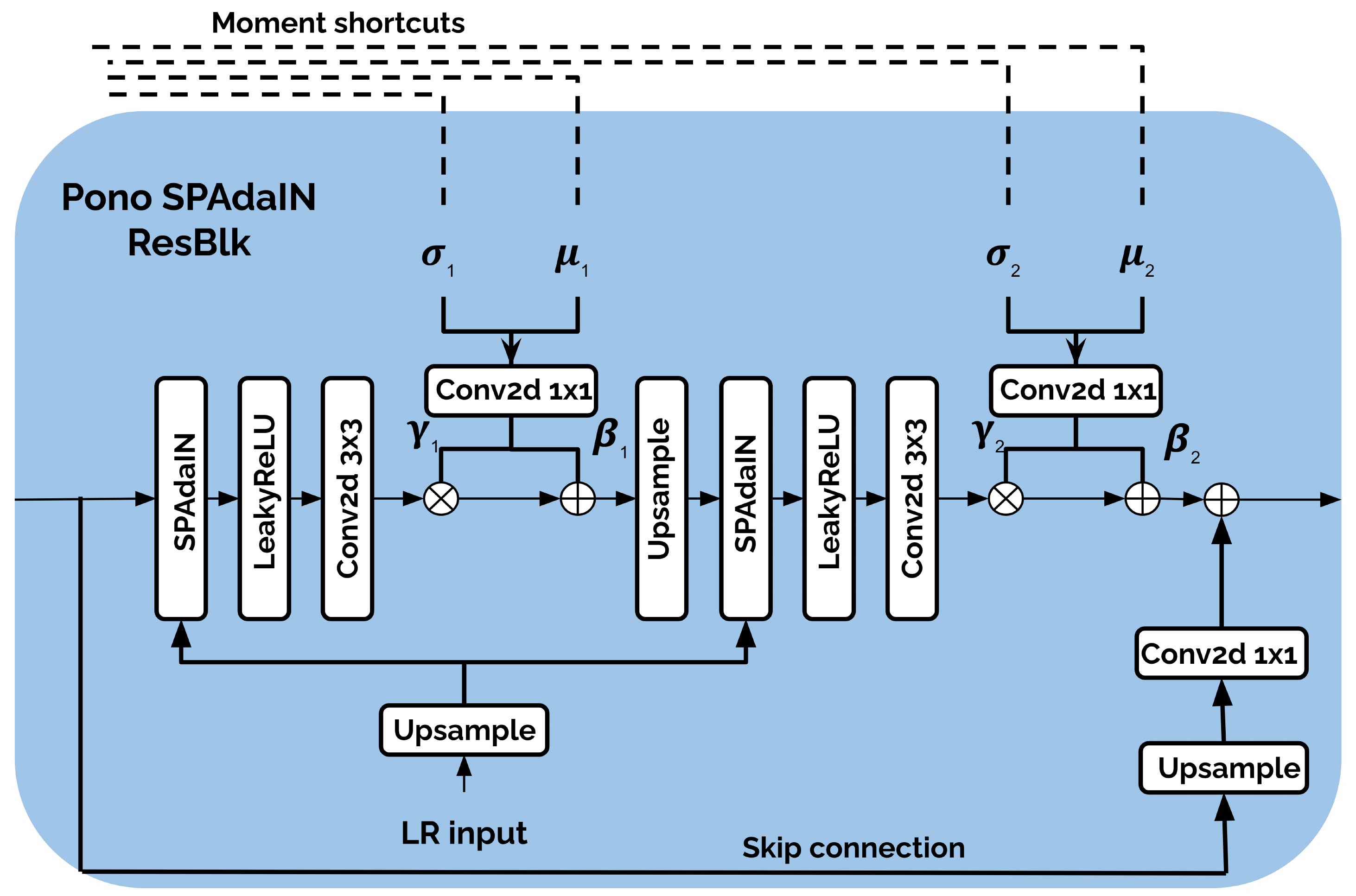}\\
    (b) Pono SPAdaIn residual blocks \\  
    \caption{Residual blocks used in the generator $G$: (a) Encoding pono residual blocks, used to extract and compress the features coming from the source image and pass it to the decoding part of the generator through moments shortcuts and as compressed features; and (b) Decoding pono SPAdaIn residual blocks, which are taking as input extracted features from the encoding part, and also are upsampling the LR target image and pass it to the SPAdaIn layer \citep{NeuralPose}.}
    \label{fig:res}
\end{figure}

Generator $G$ (\cref{fig:generator}) is U-shaped, with skip connections (a.k.a. moments shortcuts in \cite{Pono}) between the encoding and decoding part. The encoder takes the input image, passes it through a series of downsampling residual blocks (ResBlks) \citep{He_2016_CVPR}. Each ResBlks is equipped with instance normalization (IN) to remove the style of the input, followed by 2D convolution layers and a positional normalization (Pono) \citep{Pono}. The mean $\mu$ and variance $\sigma$ are subtracted and passed as a skip connection to the corresponding block in the decoder. Pono and moments shortcuts plays a crucial role in transferring the needed structural information from the source image to the decoding part of the network. These blocks, dubbed \emph{Pono ResBlks}, are illustrated in detail in \cref{fig:res} (a).

For the decoder blocks shown in \cref{fig:res}(b), we use SPAdaIN \citep{NeuralPose} conditioned on the LR image, where the LR image is first upsampled to the corresponding resolution of the \emph{Pono SPAdaIN ResBlk} using bilinear upsampling. It is then followed by 2D convolution layers and a dynamic moment shortcut layer, where, instead of reinjecting $\mu$ and $\sigma$ as is, we use a convolutional layer that takes $\mu$ and $\sigma$ as inputs to generate the $\beta$ and $\gamma$ used as moment shortcuts. Using the dynamic version of moment shortcuts \citep{Pono} allows the network to adapt and align the shape of the incoming structural information to its LR counterpart.

We use the StarGAN~v2 \citep{starganv} discriminator architecture without the domain-specific layers since we do not have predefined domains. We also concatenate the image difference (eq.~\ref{eq:ste}) at the corresponding layer (same height and width).

%!TEX root = main.tex
\section{Experiments}

\paragraph{Baseline} We compare our method with StarGAN~v2 \citep{starganv}, the state-of-the-art for image-to-image generation on CelebA-HQ and AFHQ. However, since previous i2i methods do not make use of LR images, the comparison is provided to illustrate the differences and the advantages of each method.

\paragraph{Datasets} We evaluate our method on the CelebA-HQ \citep{ProGanCelebA} and AFHQ \citep{starganv} datasets. However, for CelebA-HQ we do not separate the two domains into female and male since both domains are close to each other. Also, we are not using any extra information (e.g., facial attributes of CelebA-HQ). For AFHQ, we train our network on each domain separately (i.e., cats, dogs and wild), since the amount of information shared between these is much lower. 

\paragraph{Evaluation metrics} Baseline results are evaluated according to the metrics of image-to-image translation used in \cite{huang2018multimodal, DRIT, msgan}. Specifically, diversity and visual quality of samples produced by different methods are evaluated both with the Fr\'echet inception distance (FID) \citep{ttur} and the learned perceptual image patch similarity (LPIPS) \citep{LPIPS}.

%-------------------------------------------------------------------------
\subsection{Training Setup}

For our experiments, we fixed the LR image resolution to $8\times8$ and experimented with $128 \times 128$ and $256 \times 256$ for the HR image resolution---we ablate the effect of LR image resolution in sec.~\ref{sec:ablation}. 
We train our networks with Adam \citep{adam} and TTUR \citep{ttur}, with a learning rate of $10^{-3}$ for the generator and $4 \times 10^{-3}$ for the discriminator. We also used $R^1$ regularization \citep{Greg} with $\gamma=0.5$, with a batch size of 8. Spectral normalization \citep{miyato2018spectral} was used in all the layers of both $G$ and $D$. In \cref{Rx}, we use $\epsilon=0$ to push the downscaled version of the generated image to be as close as possible to the LR target. We set $\lambda_{cyc}=1$ when trained on $128 \times 128$, and to $\lambda_{cyc}=0.1$ for $256\times256$. 
% \todo{(CG) Does that mean that the values in Eq. 6 and 7 with a 0 distance value is different here? If so, we should revise the method section to present with a parametric $\epsilon$, not a 0.}

    \begin{figure*}
    \begin{center}
    
    \includegraphics[width=1.0\linewidth]{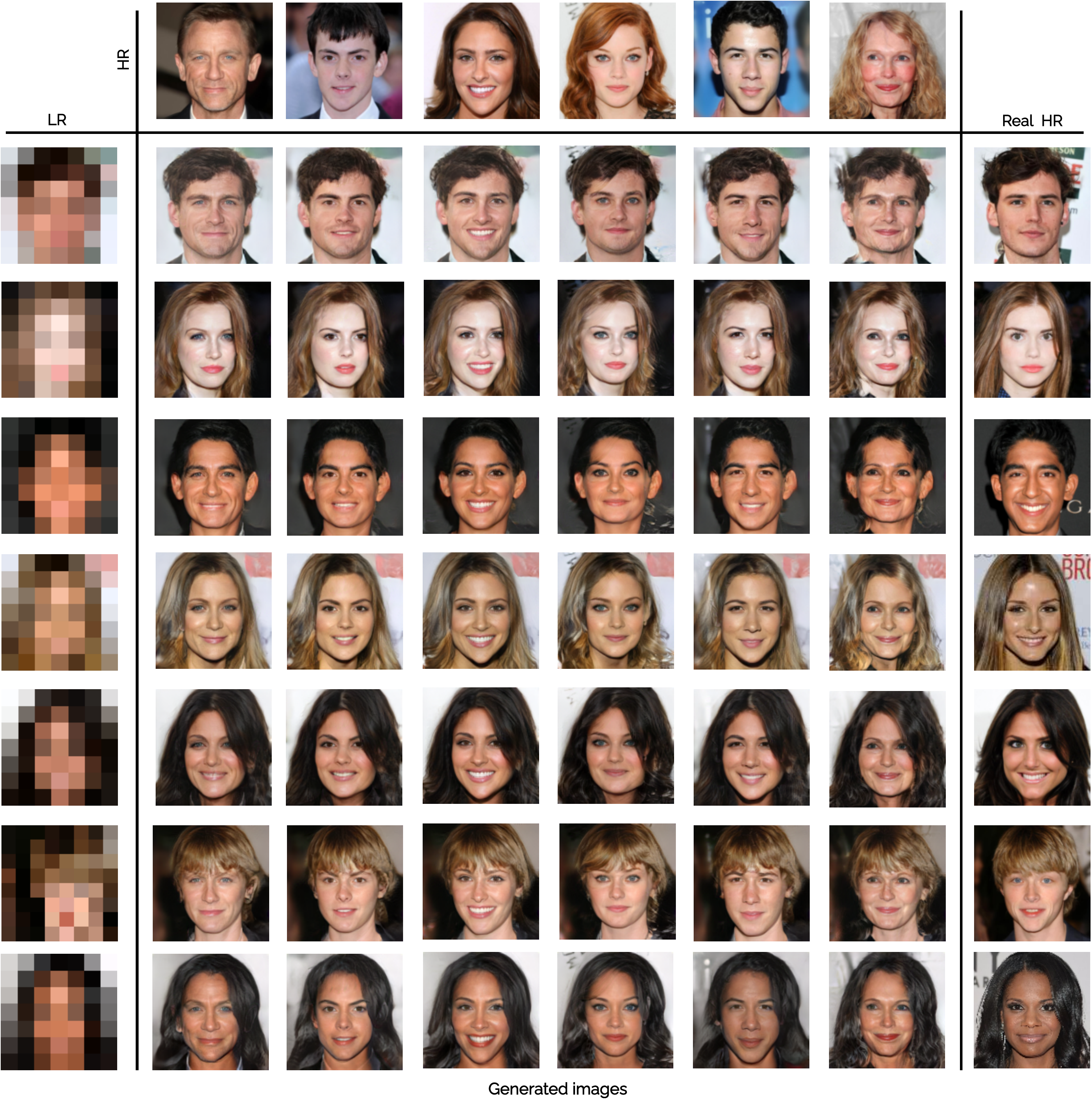}
    \end{center}
       \caption{Qualitative reference-guided image synthesis results on CelebA-HQ. Our method takes the HR source images (top row), and translates them according to the LR target (left column). We also add the real HR target (not seen by the network) for visual comparison. \emph{See supplementary material for more results}.}
    \label{fig:lots}
    \end{figure*}

\begin{figure*}
    \begin{center}
    \hspace{-1cm}
    \includegraphics[width=0.88\linewidth]{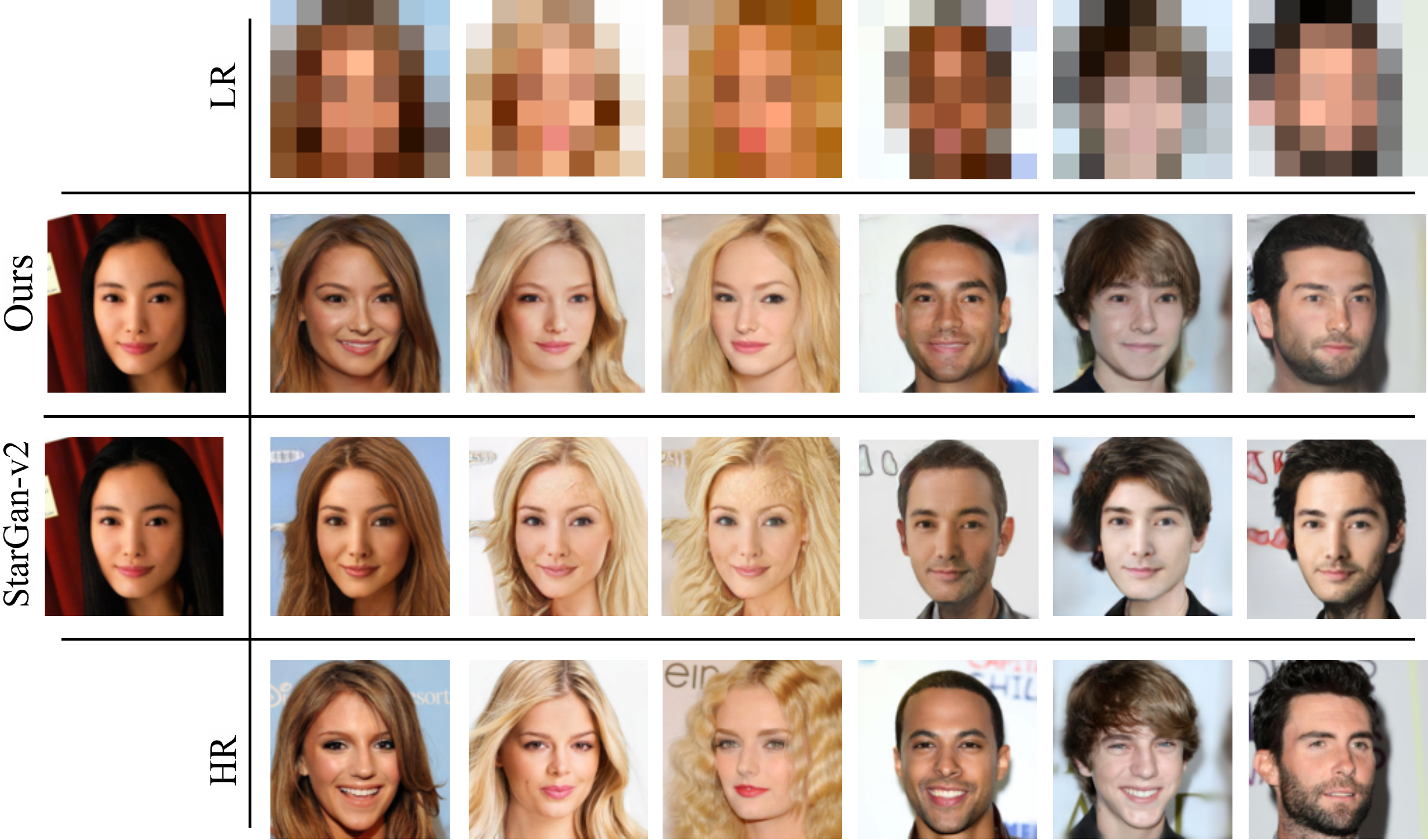}
    \end{center}
       \caption{Comparison between our method and reference guided StarGAN~v2 \citep{starganv} on CelebA-HQ. For both methods, we use the same HR image as the source (left column). For the reference image, our method uses the LR images (top row) while StarGAN~v2 \citep{starganv} uses the corresponding HR (bottom row).}
    \label{fig:comparison}
    \end{figure*}

% \begin{figure}
%     \begin{center}
%     %\hspace{-1cm}
%     \includegraphics[width=\linewidth]{fig/pulseNN.pdf}
%     \end{center}
%       \caption{Comparison between our method and PULSE~\cite{pulse} on CelebA-HQ. While both methods use the same LR image (top row), ours also leverages the reference images (left column) to guide the generation process. We also show the corresponding HR (bottom row) for reference only---neither method sees this HR image.}
%     \label{fig:pulse}
%     \end{figure}
%-------------------------------------------------------------------------
\subsection{Qualitative Evaluation}

\Cref{fig:comparison} compares images obtained with our framework with those obtained with StarGAN~v2 \citep{starganv} using reference-guided synthesis on CelebA-HQ. Since our method focuses on generating images that downscale to the given LR image, the generator learns to merge the high frequency information presents in the source image with the low frequency information of the LR target, while preserving the identity of the person and other distinctive features. Differently from traditional i2i methods that only change the style of the source image while preserving its content, our method adapts the source image to the pose of the LR target. More qualitative samples obtained with our technique are shown in \cref{fig:lots}, where the first row of HR images are used as source images and the first column is the LR target. We also display the real HR target to show that our model is capable of generating diverse images that are different from the target.

\Cref{fig:afhq} displays generated samples on AFHQ. Visually, we notice that our model is capable of merging most of the high frequency information coming from HR source image with low frequency information present in the LR target. The degree of this transfer depends on how much information is shared between the domain.  

%-------------------------------------------------------------------------
\subsection{Quantitative Evaluation}
\label{sec:results}
\begin{table}
\centering
\resizebox{\linewidth}{!}{
\begin{tabular}{lcccc}
\toprule
HR image res. & \multicolumn{2}{c}{$128\times 128$} & \multicolumn{2}{c}{$256\times 256$}  \\
Metric     & FID$_\downarrow$   & LPIPS$_\uparrow$           & FID$_\downarrow$        & LPIPS$_\uparrow$     \\ 
\midrule
MUNIT~\citep{huang2018multimodal}          & --            & --              & 107.1           & 0.176      \\
DRIT~\citep{DRIT}            & --            & --              & 53.3           & 0.311      \\
MSGAN~\citep{msgan}          & --            & --              & 39.6            & 0.312      \\
StarGAN~v2~\citep{starganv}  & 19.58         & 0.22            & \textbf{23.8}   & \textbf{0.38} \\
\midrule
Ours                         &\textbf{15.52} & \textbf{0.34}   & 25.89           & 0.329       \\
\bottomrule
\end{tabular}}
\caption{Quantitative comparison on the CelebA dataset, comparing our method to other reference-guided i2i methods \citep{starganv,huang2018multimodal, DRIT, msgan}. We follow same procedure as in StarGAN~v2 \citep{starganv}, but for our method we sample ten HR images for each LR image.}
\label{tab:metrics_fid}

\end{table}

In \cref{tab:metrics_fid}, we report FID and LPIPS scores on the results obtained on CelebA-HQ, using two different resolutions, $128\times 128$ and $256\times 256$. Results with the $256\times 256$ resolution show a significantly lower FID of our method compared to \cite{huang2018multimodal, DRIT, msgan}, while being similar to StarGAN~v2 \citep{starganv}. We notice a better FID score with the lower $128\times 128$ resolution. This is due to the fact that the task is harder with higher scale factor since we need to hallucinate more detailed textural information missing from the LR target.
\begin{figure}
\begin{center}

\includegraphics[width=0.9\linewidth]{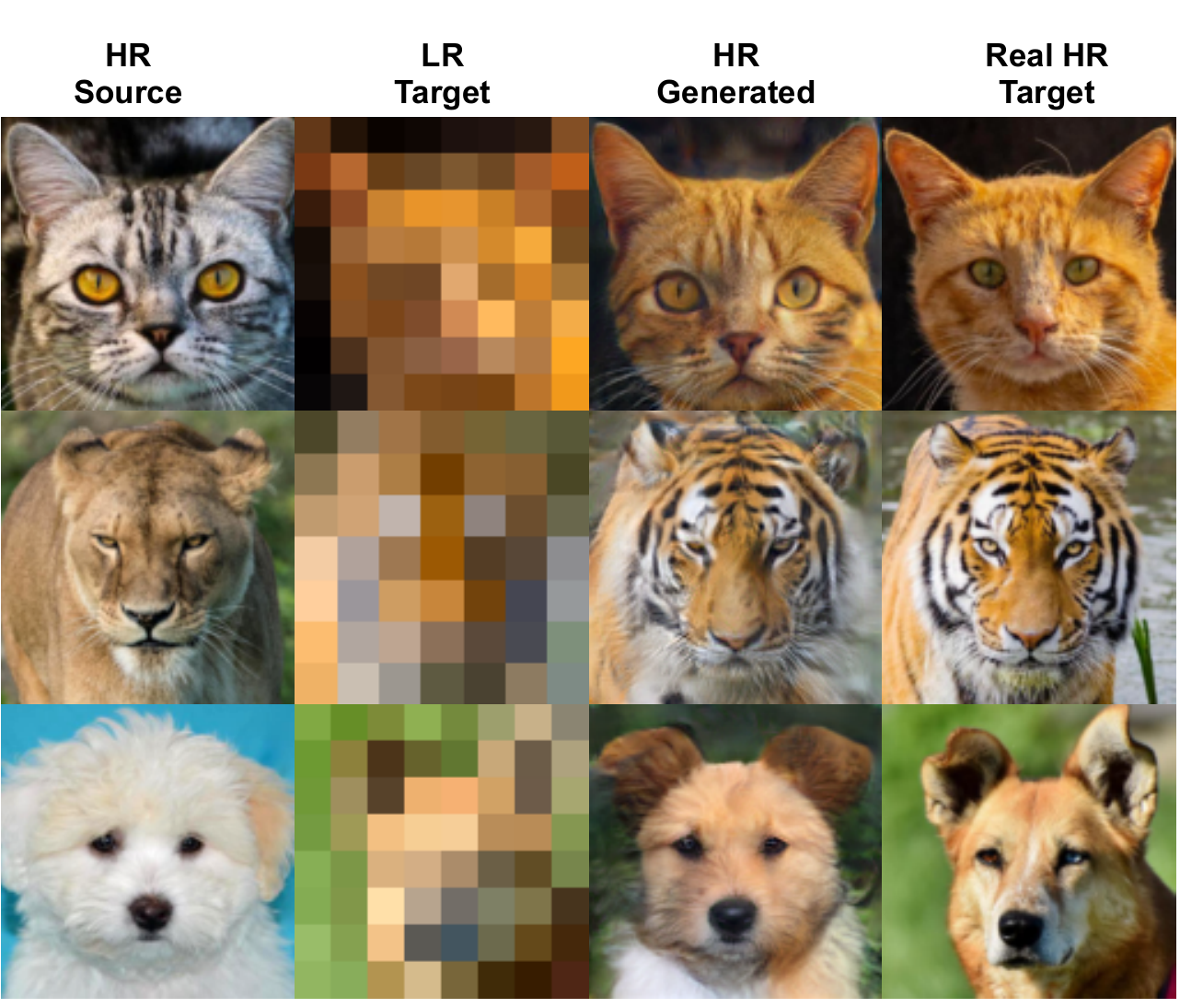}
\end{center}
   \caption{Qualitative results on the AFHQ dataset. Our method takes the HR images, and translate them to the corresponding HR subspace of the given LR target images. \emph{See supplementary material for more results}.}

\end{figure}

% For a deeper insight on the differences between our method and Stargan-v2, we used the density and coverage metrics of \cite{divCov}. The density measures the overlap between the real data and generated samples, while the coverage measures their diversity, by measuring the ratio of real samples that are covered by the fake samples~\cite{divCov}. Following \cite{divCov}, we used the feature space embedding of both the real and fakes images with a pretrained VGG16~\cite{vgg} on ImageNet. The density metric is then obtained from the $k$-nearest neighbours (with $k=5$ as in \cite{divCov}) on the 4096 features obtained from the VGG network's second fully connected layer. Diversity results reported in \cref{tab:diversity} show higher density values for our method, meaning that their samples are closer to the real data distribution than Stargan-v2. Higher coverage measures are also obtained for our method, meaning a better coverage of the data distribution modes. This is noticeable for the $128\times 128$ resolution, since the coverage is close to maximum value (the domain is $[0,1]$), being almost 10\% higher than Stargan-v2 while this gap doubles when the HR value is increased to $256\times 256$. This indicates that our model produces more realistic results by exploiting HR information from the source image, while being more diverse by staying faithful to the LR target image.

We also report quantitative results on AFHQ \citep{starganv} in \cref{tab:afhq_metrics}, where we train our model on each domain separately given the higher differences of the domains distributions. Indeed, we found that our method excels on domains where images are structurally similar and share information, such as the ``cats'' domain. However, the ``dogs'' and ``wild'' domains show a wider variety of races and species, meaning less information shared between images from the same domain. This reduces the amount of shared information that can be transferred from the HR source, forcing to hallucinate more details out of LR targets. This is confirmed by the lower LPIPS results obtained by our approach compared to StarGAN~v2, while keeping similar FID scores.

\subsection{Ablation}
\label{sec:ablation}
To investigate the robustness of our method, we report the results on the impact of resolution, the effect of grayscale LR targets and the effect of added Gaussian noise on the generation process.

\paragraph{Impact of resolution}
\Cref{tab:resolution} illustrates the impact of LR resolution on CelebA-HQ. As the LR target resolution increases, the model exploits the information in the LR target more and more over the information provided by the HR source image. This is confirmed by a sustained decrease in the LPIPS score when the resolution increases from $8\times 8$ to $32\times 32$, while maintaining a similar FID score. 

% We also test the effect of using a grayscale in the LR target images and using political faces not part of CelebA-HQ..

\paragraph{Impact of grayscale LR target}
Although our method is solely trained on RGB images, we tested our method on grayscale images (broadcasting the grayscale image 3 times along the channel axis) from CelebA-HQ dataset to investigate its impact on the SPadaIN module responsible for style guidance. We found that our model is robust to such effect and can successfully generate realistic grayscale images using HR source images. 

\begin{figure}
\centering
\includegraphics[trim=0 0.1cm 0 0, clip, width=1.\linewidth]{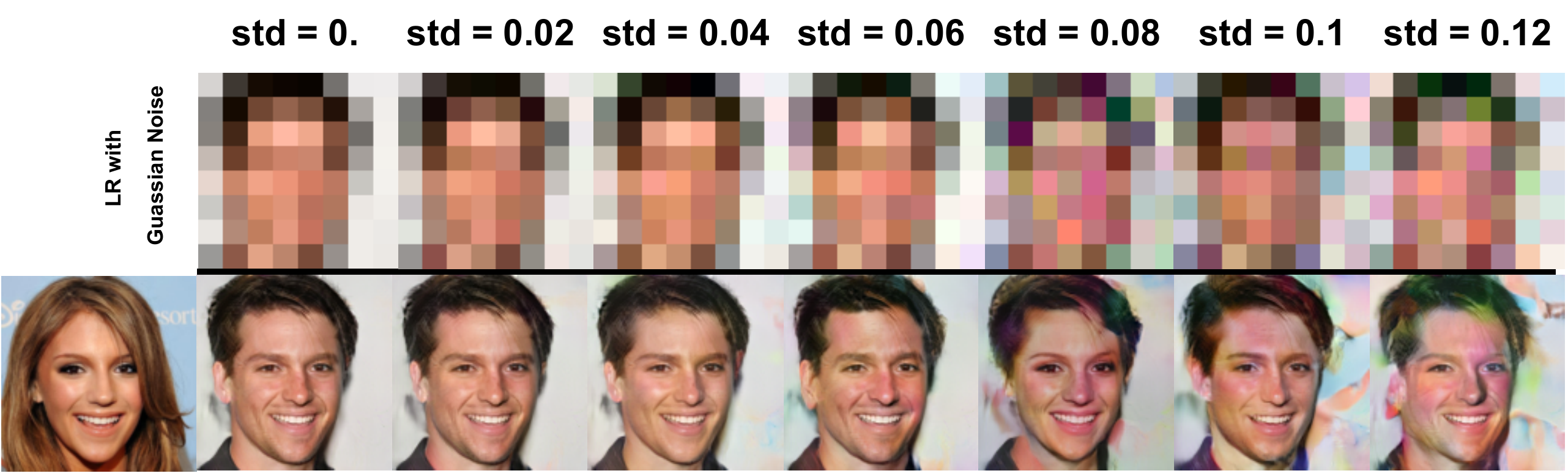}
\caption{Qualitative results with added Gaussian noise to the LR target images.}
\label{fig:noise}

\end{figure}

\paragraph{Impact of Gaussian noise}
We added Gaussian noise to the LR target image with varying $\sigma$ and found that the model is robust to $\sigma < 0.12$, see \cref{fig:noise}. Even with $\sigma > 0.10$, the model preserved well the source image features and the identity of the face.

\begin{figure}
\centering
\includegraphics[trim=0 1cm 0 0, clip, width=0.9\linewidth]{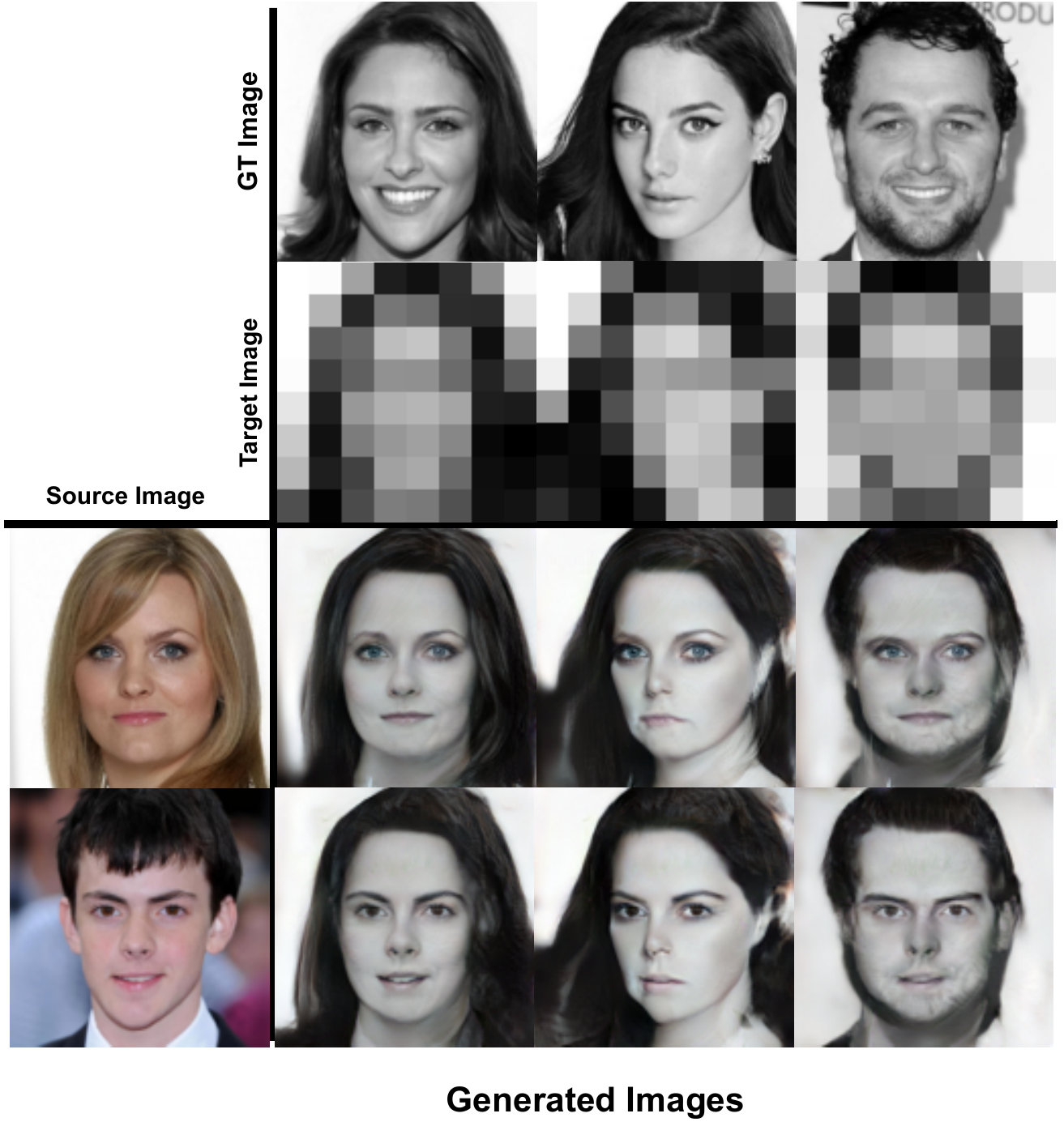}
\caption{Results of using grayscale LR target to guide the generation process.}
\label{fig:afhq}

\end{figure}

\begin{figure}
\centering
\includegraphics[trim=0 0.8cm 0 0, clip, width=1.\linewidth]{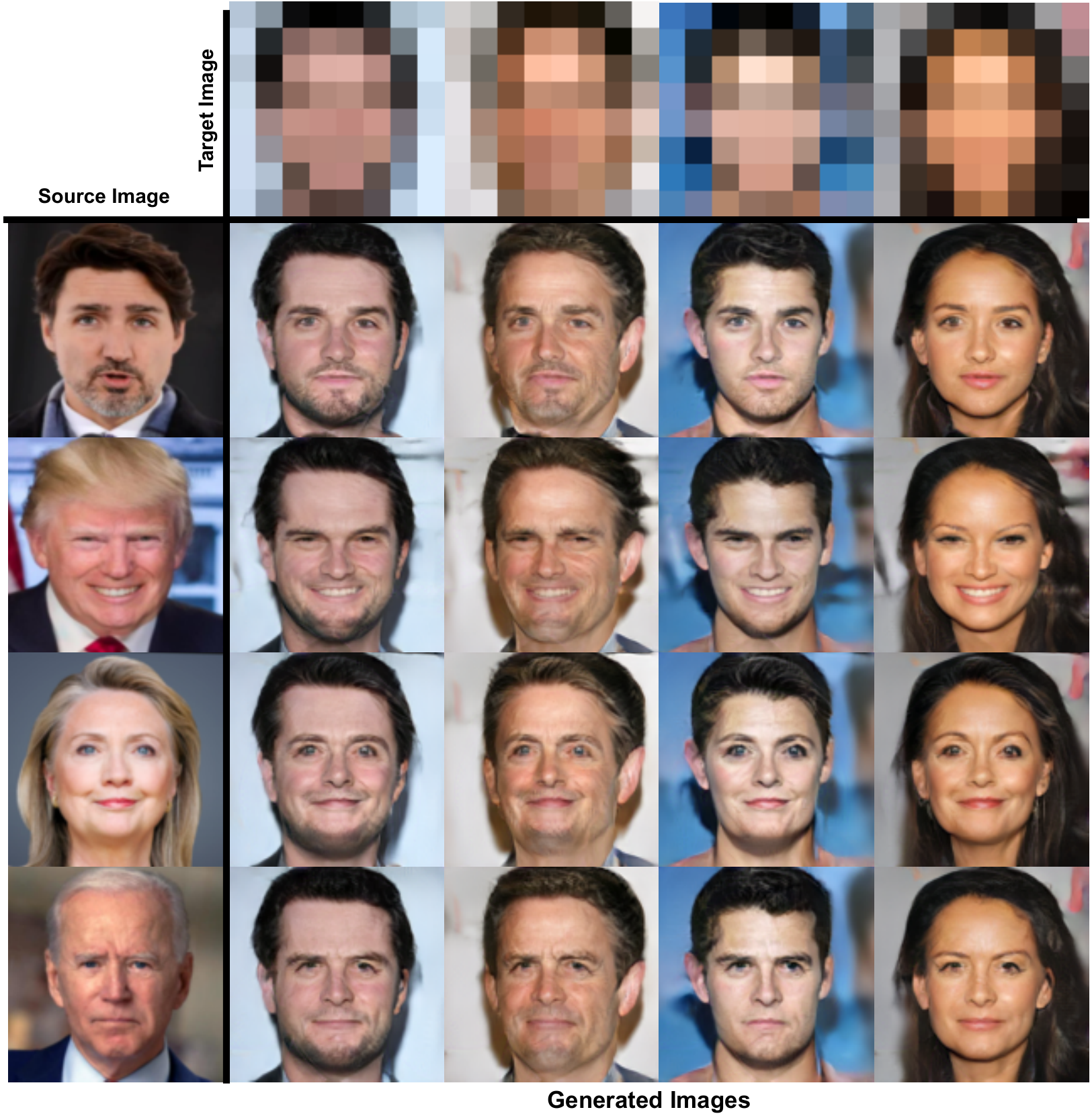}
\caption{Qualitative results on politician faces that are not part of CelebA-HQ. Our method, trained on CelebA-HQ, takes the HR images as input and translates them to the corresponding HR subspace of the given LR target images.}
\label{fig:pol}
\end{figure}

\begin{figure}
\centering
\includegraphics[trim=0 1.1cm 0 0, clip, width=0.9\linewidth]{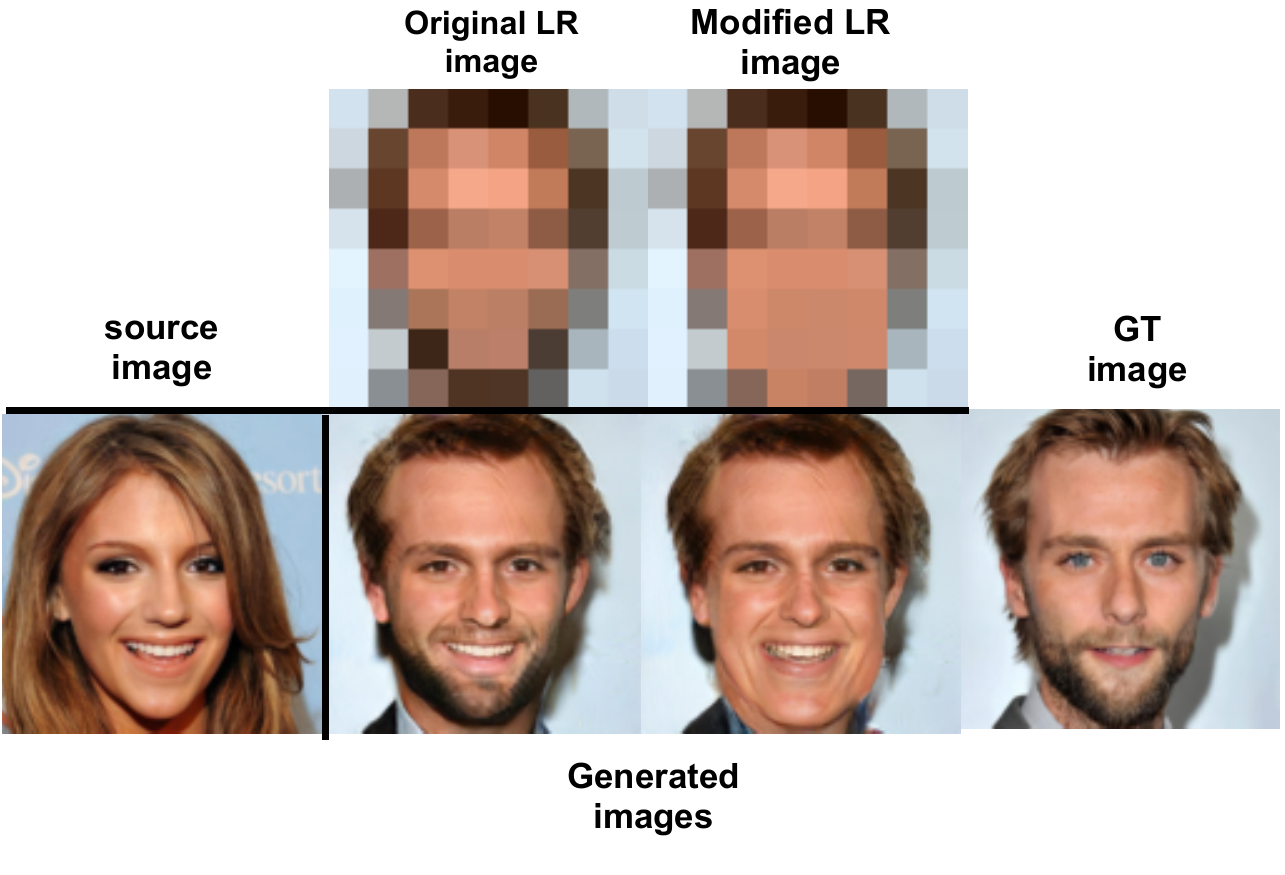}
\caption{Image generation (bottom row) can be guided by manually altering the LR target image (top row).}
\label{fig:guide}
\end{figure}

\begin{table}
\centering
\resizebox{\linewidth}{!}{%
\begin{tabular}{lcccccc}
\toprule
Category & \multicolumn{2}{c}{Cats}     & \multicolumn{2}{c}{Dogs}             & \multicolumn{2}{c}{Wild} \\ 
Metric & FID$_\downarrow$  & LPIPS$_\uparrow$     & FID$_\downarrow$   & LPIPS$_\uparrow$     & FID$_\downarrow$         & LPIPS$_\uparrow$       \\ 
\midrule
StarGAN~v2  & 25.2 & \textbf{0.42} & \textbf{56.5} & 0.34 & \textbf{19.87} & \textbf{0.46} \\
Ours        & \textbf{22.8} & 0.40      & 67.3 & \textbf{0.47}      & 20.61       & 0.23        \\ 
\bottomrule
\end{tabular}%
}
\caption{Quantitative comparison on the AFHQ dataset. We compare our method to the reference-guided i2i methods StarGAN~v2 \citep{starganv} where the target and the source image are from the same domain.}
\label{tab:afhq_metrics}

\end{table}

\begin{table}[]
\centering
% \resizebox{\linewidth}{!}{%
\begin{tabular}{lcccc}
\toprule
LR target res. & $4\times 4$ & $8\times 8$        & $16\times 16$       & $32\times 32$       \\ 
\midrule
FID$_\downarrow$        & 15.13             & 15.34    & 19.45    & 13.55    \\
LPIPS$_\uparrow$        & 0.30           & 0.34     & 0.14     & 0.08     \\ 
\bottomrule
\end{tabular}%
% }
\caption{Effect of the LR resolution on the FID and LPIPS metrics for the CelebA-HQ dataset.}
\label{tab:resolution}

\end{table}

\subsection{Manual guidance}
We show that our method can allow for manual controlled generation with fairly minimal changes in the LR target image. \cref{fig:guide} shows that manually altering pixel values in the LR image (beard) can generate a realistic change in the output. %Such guidance can be intuitive to the user since the generation is aligned with the user visual perspective.  

\subsection{StarGAN~v2 under same setup}
\begin{figure}
\centering
\includegraphics[width=0.8\linewidth]{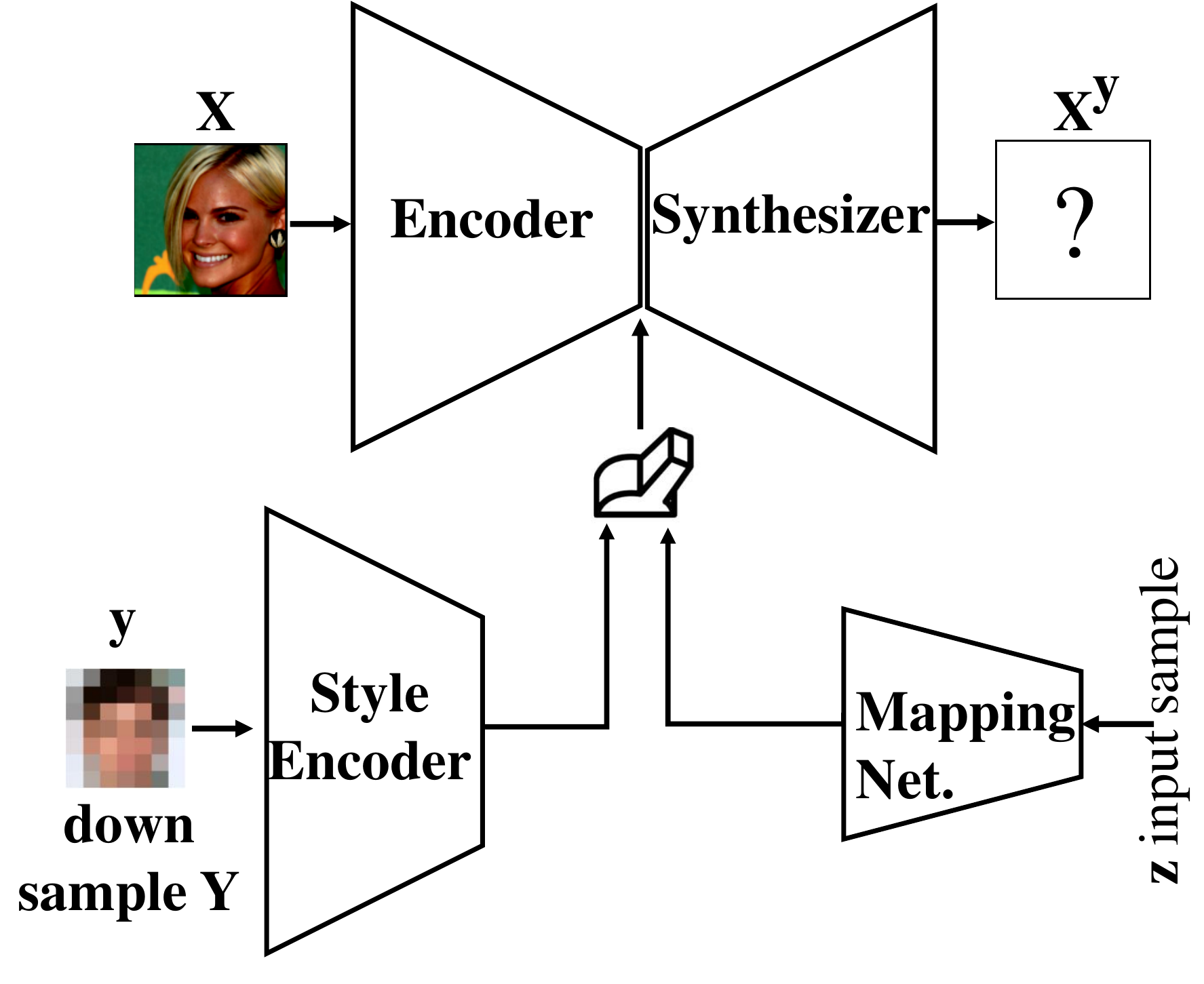}
\caption{StarGAN~v2 conditioned on LR. We ablate StarGAN~v2 under two setups with and without a mapping network while the style encoder takes the LR input image.} %\todo{That figure is really not clear and I am not sure what we what to carry in it, we need to reconsider it.}}
\label{fig:stargan_analysis}
\end{figure}

\begin{figure}
\centering
\includegraphics[width=0.99\linewidth]{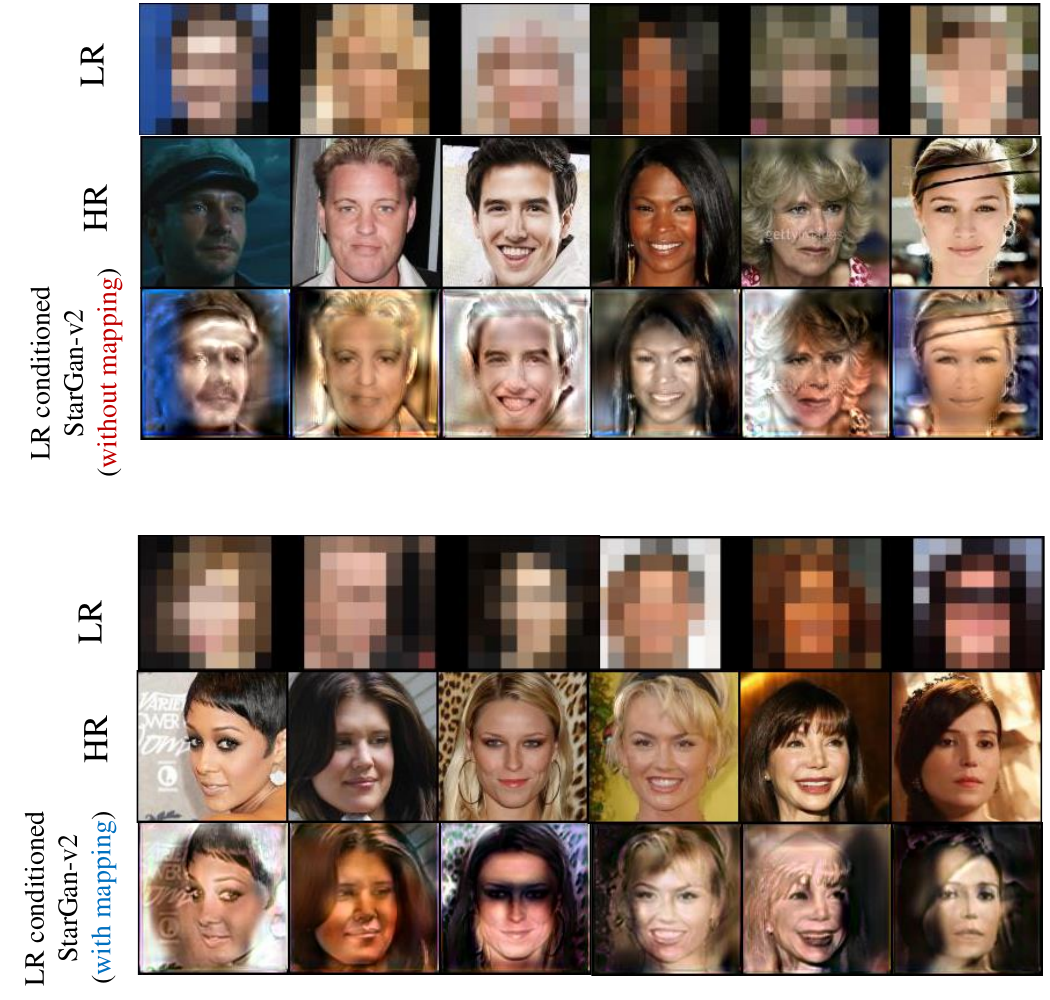}
\caption{Examining the qualitative outcomes of conditioning StarGAN v2 using low-resolution images, in order to assess the proficiency of StarGAN v2 within the context of our domain-agnostic image-to-image translation.  Here, we consider two versions of StarGan v2, without a mapping network (top three rows) and with a mapping network (down three rows).}
\label{fig:conditioned_starganv2}
\end{figure}

Unlike StarGAN~v2 \citep{starganv} with two additional learning modules besides generator and discriminator, our proposed model is domain-agnostic. Nevertheless, StarGAN~v2 is still the most related work to our proposed model from i2i perspective. Therefore, we perform an ablative analysis of StarGAN~v2 under the LR conditioning hypothesis, as shown in \cref{fig:stargan_analysis}. Particularly, we train the model by LR images as the input of the Style encoder. First, we kept the original architecture of StarGAN~v2 and only conditioned the input of the style network on LR input image. Second, we train StarGAN~v2 with LR inputs to style the network and without the mapping network to reduce one of the extra modules used in the original paper. As depicted in fig.~\cref{fig:conditioned_starganv2}, StarGAN v2 struggles to produce realistic images when subjected to conditions akin to our domain-agnostic image-to-image translation.

% \begin{figure} 
%     \centering
%     \footnotesize
%     \setlength{\tabcolsep}{1pt}
%     \begin{tabular}{ccc} 
%     \includegraphics[width=0.48\linewidth, angle=0]{fig_stargan/StarGanFig1.pdf} & 
%     \includegraphics[width=0.48\linewidth, angle=0]{fig_stargan/StarGanFig2.pdf} &   
%     \\ \ \\ 
%     (a) $\mathbf{X}$ in the style of  $\mathbf{Y}$ & 
%     (b) $\mathbf{Y}$ in the style of  $\mathbf{X}$ \\  
%     \end{tabular}
%     % \vspace{0.5em}
%     \caption{Illustration of our domain-agnostic method. The Generator $ G_{\theta}(\cdot, \cdot)$ is trained to map a HR  $\mathbf{Y}$ (resp. $\mathbf{X}$) from its domain $\mathcal{M}_\mathbf{Y}$ (resp. $\mathcal{M}_\mathbf{X}$) to the other domain $\mathcal{M}_\mathbf{X}$ (resp. $\mathcal{M}_\mathbf{Y}$), conditioned on a LR version $\mathbf{x}$ of $\mathbf{X}$ (resp. $\mathbf{y}$ of $\mathbf{Y}$). 
%     a) Translation of HR image $\mathbf{Y}$ from $\mathcal{M}_\mathbf{Y}$ to $\mathcal{M}_\mathbf{X}$ guided by LR target $\mathbf{x}$.
%     b) Translation of HR image $\mathbf{X}$ from $\mathcal{M}_\mathbf{X}$ to $\mathcal{M}_\mathbf{Y}$ guided by LR target $\mathbf{y}$.
%     }% 
%     \label{fig:space_illustration}
%     % \label{fig:idea_illustration}
% \end{figure}

%!TEX root = main.tex
\section{Discussion}

% Order: +/-/+
% Start by summarizing work, highlighting contributions
% Limitations and future work 

This paper proposes a novel framework for reference-guided image synthesis. Our method attempts to realistically project the source images to a subspace of images that shares same features with a low resolution target (such as overall color distribution and pose). Our experiments show that our method allows for the generation of a wide variety of realistic images given LR targets. We validate our method on two well-known datasets, CelebA-HQ and AFHQ and we present its difference and advantages compared to the leading i2i methods \cite{starganv, huang2018multimodal, DRIT}.  We also investigated the robustness of our method in terms of flexibility using grayscale images and its generalization on new faces outside of the dataset it is trained on. We also show that our method allows for a manual guidance with fairly minimal changes, such guidance can be intuitive and can be made directly on the LR target. 

\subsection{Limitations and future work}
The main limitation of our method is that it assumes the target and source images come from the same distribution (e.g., human faces). In the case where these domains differ (e.g., tiger vs lion in the wild category of AFHQ), the generated image attempts to match the LR target at the expense of ``forgetting'' more information from the source image (\cref{fig:afhq}). % In addition, we also note that results sometimes lack diversity for a given LR target (rows in \cref{fig:lots}, for example)---this is a consequence of having to perfectly match the LR image.
A potential solution to mitigate this problem would be to soften the constraint of having to perfectly match the LR image, for instance by increasing the distance $\epsilon$ in \cref{Rx}, or by modifying the discriminator inputs (\cref{eq:disInput}) to tolerate larger differences. Finally, the framework has so far only been tested on faces (humans and animals). Extending it to handle more generic scenes, where the LR target would capture higher level information such as layout, makes for an exciting future research direction. 

% on closely related images and on very low resolution targets, where it can excel by generating high quality, realistic images. 
% Although, it's possible to apply on domains containing wide varieties, it can hurt the diversity of the model and reduce the similarities between the source image and the generated one, which can be considered as the main limitation of our work. Such issues will be the subject of our future works.   

% \input{fig_starganV2} 

\section*{Acknowledgements}

This work was done in the context of a collaborative R\&D project with Thales Canada, which provided expertise, resources, support and funding. The project has been funded by the Natural Sciences and Engineering Research Council of Canada (NSERC) (RDCPJ 537836-18).
 
\bibliography{main_arxiv}

\begin{thebibliography}{45}
\providecommand{\natexlab}[1]{#1}
\providecommand{\url}[1]{\texttt{#1}}
\expandafter\ifx\csname urlstyle\endcsname\relax
  \providecommand{\doi}[1]{doi: #1}\else
  \providecommand{\doi}{doi: \begingroup \urlstyle{rm}\Url}\fi

\bibitem[Arjovsky et~al.(2017)Arjovsky, Chintala, and Bottou]{Wgan}
Martin Arjovsky, Soumith Chintala, and L\'eon Bottou.
\newblock {W}asserstein generative adversarial networks.
\newblock In \emph{International Conference on Machine Learning (ICML)}, 2017.
\newblock URL \url{https://proceedings.mlr.press/v70/arjovsky17a.html}.

\bibitem[Armanious et~al.(2020)Armanious, Jiang, Fischer, K{\"u}stner, Hepp,
  Nikolaou, Gatidis, and Yang]{armanious2020medgan}
Karim Armanious, Chenming Jiang, Marc Fischer, Thomas K{\"u}stner, Tobias Hepp,
  Konstantin Nikolaou, Sergios Gatidis, and Bin Yang.
\newblock {MedGAN}: Medical image translation using {GANs}.
\newblock \emph{Computerized medical imaging and graphics}, 79:\penalty0
  101684, 2020.
\newblock URL \url{https://doi.org/10.1016/j.compmedimag.2019.101684}.

\bibitem[Bengio et~al.(2013)Bengio, L\'{e}onard, and Courville]{steEstimator}
Yoshua Bengio, Nicholas L\'{e}onard, and Aaron~C. Courville.
\newblock Estimating or propagating gradients through stochastic neurons for
  conditional computation.
\newblock \emph{arXiv preprint arXiv:1308.3432}, 2013.
\newblock URL \url{https://arxiv.org/abs/1308.3432}.

\bibitem[Berthelot et~al.(2020)Berthelot, Milanfar, and
  Goodfellow]{berthelot2020creating}
David Berthelot, Peyman Milanfar, and Ian Goodfellow.
\newblock Creating high resolution images with a latent adversarial generator.
\newblock \emph{arXiv preprint arXiv:2003.02365}, 2020.
\newblock URL \url{https://arxiv.org/abs/2003.02365}.

\bibitem[Brock et~al.(2018)Brock, Donahue, and Simonyan]{Biggan}
Andrew Brock, Jeff Donahue, and Karen Simonyan.
\newblock Large scale {GAN} training for high fidelity natural image synthesis.
\newblock In \emph{International Conference on Learning Representations
  (ICLR)}, 2018.
\newblock URL \url{https://arxiv.org/abs/1809.11096}.

\bibitem[Choi et~al.(2018)Choi, Choi, Kim, Ha, Kim, and Choo]{stargan}
Yunjey Choi, Minje Choi, Munyoung Kim, Jung-Woo Ha, Sunghun Kim, and Jaegul
  Choo.
\newblock {StarGAN}: Unified generative adversarial networks for multi-domain
  image-to-image translation.
\newblock In \emph{IEEE Conference on Computer Vision and Pattern Recognition
  (CVPR)}, 2018.
\newblock URL
  \url{https://openaccess.thecvf.com/content_cvpr_2018/html/Choi_StarGAN_Unified_Generative_CVPR_2018_paper.html}.

\bibitem[Choi et~al.(2020)Choi, Uh, Yoo, and Ha]{starganv}
Yunjey Choi, Youngjung Uh, Jaejun Yoo, and Jung-Woo Ha.
\newblock {StarGAN v2}: Diverse image synthesis for multiple domains.
\newblock In \emph{{IEEE/CVF} Conference on Computer Vision and Pattern
  Recognition (CVPR)}, 2020.
\newblock URL
  \url{https://openaccess.thecvf.com/content_CVPR_2020/html/Choi_StarGAN_v2_Diverse_Image_Synthesis_for_Multiple_Domains_CVPR_2020_paper.html}.

\bibitem[Dong et~al.(2014)Dong, Loy, He, and Tang]{dong2014learning}
Chao Dong, Chen~Change Loy, Kaiming He, and Xiaoou Tang.
\newblock Learning a deep convolutional network for image super-resolution.
\newblock In \emph{European Conference on Computer Vision (ECCV)}, 2014.
\newblock URL \url{https://doi.org/10.1007/978-3-319-10593-2_13}.

\bibitem[Goodfellow et~al.(2014)Goodfellow, Pouget-Abadie, Mirza, Xu,
  Warde-Farley, Ozair, Courville, and Bengio]{goodfellow}
Ian Goodfellow, Jean Pouget-Abadie, Mehdi Mirza, Bing Xu, David Warde-Farley,
  Sherjil Ozair, Aaron Courville, and Yoshua Bengio.
\newblock Generative adversarial nets.
\newblock In \emph{Advances in Neural Information Processing Systems (NIPS)},
  2014.
\newblock URL \url{https://arxiv.org/abs/1406.2661}.

\bibitem[Goodfellow et~al.(2020)Goodfellow, Pouget-Abadie, Mirza, Xu,
  Warde-Farley, Ozair, Courville, and Bengio]{goodfellow2020generative}
Ian Goodfellow, Jean Pouget-Abadie, Mehdi Mirza, Bing Xu, David Warde-Farley,
  Sherjil Ozair, Aaron Courville, and Yoshua Bengio.
\newblock Generative adversarial networks.
\newblock \emph{Communications of the ACM}, 63\penalty0 (11):\penalty0
  139--144, 2020.
\newblock URL \url{https://dl.acm.org/doi/abs/10.1145/3422622}.

\bibitem[Gulrajani et~al.(2017)Gulrajani, Ahmed, Arjovsky, Dumoulin, and
  Courville]{wgangp}
Ishaan Gulrajani, Faruk Ahmed, Martin Arjovsky, Vincent Dumoulin, and Aaron~C
  Courville.
\newblock Improved training of {W}asserstein {GANs}.
\newblock In \emph{Advances in Neural Information Processing Systems
  (NeurIPS)}, 2017.
\newblock URL
  \url{https://proceedings.neurips.cc/paper_files/paper/2017/hash/892c3b1c6dccd52936e27cbd0ff683d6-Abstract.html}.

\bibitem[He et~al.(2016)He, Zhang, Ren, and Sun]{He_2016_CVPR}
Kaiming He, Xiangyu Zhang, Shaoqing Ren, and Jian Sun.
\newblock Deep residual learning for image recognition.
\newblock In \emph{IEEE Conference on Computer Vision and Pattern Recognition
  (CVPR)}, 2016.
\newblock URL
  \url{https://openaccess.thecvf.com/content_cvpr_2016/html/He_Deep_Residual_Learning_CVPR_2016_paper.html}.

\bibitem[Heusel et~al.(2017)Heusel, Ramsauer, Unterthiner, Nessler, and
  Hochreiter]{ttur}
Martin Heusel, Hubert Ramsauer, Thomas Unterthiner, Bernhard Nessler, and Sepp
  Hochreiter.
\newblock {GANs} trained by a two time-scale update rule converge to a local
  nash equilibrium.
\newblock In \emph{Advances in Neural Information Processing Systems
  (NeurIPS)}, 2017.
\newblock URL
  \url{https://proceedings.neurips.cc/paper_files/paper/2017/hash/8a1d694707eb0fefe65871369074926d-Abstract.html}.

\bibitem[Hoyer et~al.(2022)Hoyer, Dai, and Van~Gool]{hoyer2022hrda}
Lukas Hoyer, Dengxin Dai, and Luc Van~Gool.
\newblock Hrda: Context-aware high-resolution domain-adaptive semantic
  segmentation.
\newblock In \emph{European Conference on Computer Vision}. Springer, 2022.

\bibitem[Hu et~al.(2022)Hu, Huang, Shi, Li, Gao, Sun, and Li]{hu2022style}
Xueqi Hu, Qiusheng Huang, Zhengyi Shi, Siyuan Li, Changxin Gao, Li~Sun, and
  Qingli Li.
\newblock Style transformer for image inversion and editing.
\newblock In \emph{Conference on Computer Vision and Pattern Recognition},
  2022.

\bibitem[Huang and Belongie(2017)]{huang2017arbitrary}
Xun Huang and Serge Belongie.
\newblock Arbitrary style transfer in real-time with adaptive instance
  normalization.
\newblock In \emph{{IEEE} International Conference on Computer Vision (ICCV)},
  2017.
\newblock URL
  \url{https://openaccess.thecvf.com/content_iccv_2017/html/Huang_Arbitrary_Style_Transfer_ICCV_2017_paper.html}.

\bibitem[Huang et~al.(2018)Huang, Liu, Belongie, and
  Kautz]{huang2018multimodal}
Xun Huang, Ming-Yu Liu, Serge Belongie, and Jan Kautz.
\newblock Multimodal unsupervised image-to-image translation.
\newblock In \emph{European Conference on Computer Vision (ECCV)}, 2018.
\newblock URL
  \url{https://openaccess.thecvf.com/content_ECCV_2018/html/Xun_Huang_Multimodal_Unsupervised_Image-to-image_ECCV_2018_paper.html}.

\bibitem[Isola et~al.(2017)Isola, Zhu, Zhou, and Efros]{isola2017image}
Phillip Isola, Jun-Yan Zhu, Tinghui Zhou, and Alexei~A Efros.
\newblock Image-to-image translation with conditional adversarial networks.
\newblock In \emph{IEEE Conference on Computer Vision and Pattern Recognition
  (CVPR)}, 2017.
\newblock URL
  \url{https://openaccess.thecvf.com/content_cvpr_2017/html/Isola_Image-To-Image_Translation_With_CVPR_2017_paper.html}.

\bibitem[Karras et~al.(2018)Karras, Aila, Laine, and Lehtinen]{ProGanCelebA}
Tero Karras, Timo Aila, Samuli Laine, and Jaakko Lehtinen.
\newblock Progressive growing of {GAN}s for improved quality, stability, and
  variation.
\newblock In \emph{International Conference on Learning Representations
  (ICLR)}, 2018.
\newblock URL \url{https://arxiv.org/abs/1710.10196}.

\bibitem[Karras et~al.(2019)Karras, Laine, and Aila]{karras2019style}
Tero Karras, Samuli Laine, and Timo Aila.
\newblock A style-based generator architecture for generative adversarial
  networks.
\newblock In \emph{IEEE/CVF Conference on Computer Vision and Pattern
  Recognition (CVPR)}, 2019.
\newblock URL
  \url{https://openaccess.thecvf.com/content_CVPR_2019/html/Karras_A_Style-Based_Generator_Architecture_for_Generative_Adversarial_Networks_CVPR_2019_paper.html}.

\bibitem[Karras et~al.(2020)Karras, Laine, Aittala, Hellsten, Lehtinen, and
  Aila]{stylegan2}
Tero Karras, Samuli Laine, Miika Aittala, Janne Hellsten, Jaakko Lehtinen, and
  Timo Aila.
\newblock Analyzing and improving the image quality of {StyleGAN}.
\newblock In \emph{IEEE/CVF Conference on Computer Vision and Pattern
  Recognition (CVPR)}, 2020.
\newblock URL
  \url{https://openaccess.thecvf.com/content_CVPR_2020/html/Karras_Analyzing_and_Improving_the_Image_Quality_of_StyleGAN_CVPR_2020_paper.html}.

\bibitem[Kim et~al.(2019)Kim, Kim, Kang, and Lee]{kim2019u}
Junho Kim, Minjae Kim, Hyeonwoo Kang, and Kwanghee Lee.
\newblock U-gat-it: Unsupervised generative attentional networks with adaptive
  layer-instance normalization for image-to-image translation.
\newblock \emph{arXiv preprint arXiv:1907.10830}, 2019.

\bibitem[Kim et~al.(2022{\natexlab{a}})Kim, Baek, Park, Kim, and
  Kim]{kim2022instaformer}
Soohyun Kim, Jongbeom Baek, Jihye Park, Gyeongnyeon Kim, and Seungryong Kim.
\newblock Instaformer: Instance-aware image-to-image translation with
  transformer.
\newblock In \emph{Conference on Computer Vision and Pattern Recognition},
  2022{\natexlab{a}}.

\bibitem[Kim et~al.(2022{\natexlab{b}})Kim, Kang, Shin, Yoon, Eom, Shin, and
  Yun]{kim2022region}
Taehyeon Kim, Shinhwan Kang, Hyeonjeong Shin, Deukryeol Yoon, Seongha Eom,
  Kijung Shin, and Se-Young Yun.
\newblock Region-conditioned orthogonal 3d u-net for weather4cast competition.
\newblock \emph{Conference on Neural Information Processing Systems},
  2022{\natexlab{b}}.

\bibitem[Kingma and Ba(2015)]{adam}
Diederik~P. Kingma and Jimmy Ba.
\newblock Adam: {A} method for stochastic optimization.
\newblock In \emph{International Conference on Learning Representations
  (ICLR)}, 2015.
\newblock URL \url{https://arxiv.org/abs/1412.6980}.

\bibitem[Ledig et~al.(2017)Ledig, Theis, Husz{\'a}r, Caballero, Cunningham,
  Acosta, Aitken, Tejani, Totz, Wang, et~al.]{srgan}
Christian Ledig, Lucas Theis, Ferenc Husz{\'a}r, Jose Caballero, Andrew
  Cunningham, Alejandro Acosta, Andrew Aitken, Alykhan Tejani, Johannes Totz,
  Zehan Wang, et~al.
\newblock Photo-realistic single image super-resolution using a generative
  adversarial network.
\newblock In \emph{IEEE Conference on Computer Vision and Pattern Recognition
  (CVPR)}, 2017.
\newblock URL
  \url{https://openaccess.thecvf.com/content_cvpr_2017/html/Ledig_Photo-Realistic_Single_Image_CVPR_2017_paper.html}.

\bibitem[Lee et~al.(2018)Lee, Tseng, Huang, Singh, and Yang]{DRIT}
Hsin-Ying Lee, Hung-Yu Tseng, Jia-Bin Huang, Maneesh Singh, and Ming-Hsuan
  Yang.
\newblock Diverse image-to-image translation via disentangled representations.
\newblock In \emph{European Conference on Computer Vision (ECCV)}, 2018.
\newblock URL
  \url{https://openaccess.thecvf.com/content_ECCV_2018/html/Hsin-Ying_Lee_Diverse_Image-to-Image_Translation_ECCV_2018_paper.html}.

\bibitem[Li et~al.(2019)Li, Wu, Weinberger, and Belongie]{Pono}
Boyi Li, Felix Wu, Kilian~Q Weinberger, and Serge Belongie.
\newblock Positional normalization.
\newblock In \emph{Advances in Neural Information Processing Systems
  (NeurIPS)}, 2019.
\newblock URL
  \url{https://proceedings.neurips.cc/paper/2019/hash/6d0f846348a856321729a2f36734d1a7-Abstract.html}.

\bibitem[Lin et~al.(2018)Lin, Xia, Qin, Chen, and Liu]{CItoI}
Jianxin Lin, Yingce Xia, Tao Qin, Zhibo Chen, and Tie-Yan Liu.
\newblock Conditional image-to-image translation.
\newblock In \emph{IEEE Conference on Computer Vision and Pattern Recognition
  (CVPR)}, 2018.
\newblock URL
  \url{https://openaccess.thecvf.com/content_cvpr_2018/html/Lin_Conditional_Image-to-Image_Translation_CVPR_2018_paper.html}.

\bibitem[Liu et~al.(2017)Liu, Breuel, and Kautz]{liu2017unsupervised}
Ming-Yu Liu, Thomas Breuel, and Jan Kautz.
\newblock Unsupervised image-to-image translation networks.
\newblock In \emph{Advances in neural information processing systems (NIPS)},
  2017.
\newblock URL
  \url{https://proceedings.neurips.cc/paper_files/paper/2017/hash/dc6a6489640ca02b0d42dabeb8e46bb7-Abstract.html}.

\bibitem[Mao et~al.(2019)Mao, Lee, Tseng, Ma, and Yang]{msgan}
Qi~Mao, Hsin-Ying Lee, Hung-Yu Tseng, Siwei Ma, and Ming-Hsuan Yang.
\newblock Mode seeking generative adversarial networks for diverse image
  synthesis.
\newblock In \emph{IEEE/CVF Conference on Computer Vision and Pattern
  Recognition (CVPR)}, 2019.
\newblock URL
  \url{https://openaccess.thecvf.com/content_CVPR_2019/html/Mao_Mode_Seeking_Generative_Adversarial_Networks_for_Diverse_Image_Synthesis_CVPR_2019_paper.html}.

\bibitem[Mescheder et~al.(2018)Mescheder, Nowozin, and Geiger]{Greg}
Lars Mescheder, Sebastian Nowozin, and Andreas Geiger.
\newblock Which training methods for {GANs} do actually converge?
\newblock In \emph{International Conference on Machine Learning (ICML)}, 2018.
\newblock URL \url{https://proceedings.mlr.press/v80/mescheder18a}.

\bibitem[Miyato et~al.(2018)Miyato, Kataoka, Koyama, and
  Yoshida]{miyato2018spectral}
Takeru Miyato, Toshiki Kataoka, Masanori Koyama, and Yuichi Yoshida.
\newblock Spectral normalization for generative adversarial networks.
\newblock In \emph{International Conference on Learning Representations
  (ICLR)}, 2018.
\newblock URL \url{https://arxiv.org/abs/1802.05957}.

\bibitem[Su et~al.(2020)Su, Chu, and Huang]{color}
Jheng-Wei Su, Hung-Kuo Chu, and Jia-Bin Huang.
\newblock Instance-aware image colorization.
\newblock In \emph{IEEE/CVF Conference on Computer Vision and Pattern
  Recognition (CVPR)}, 2020.
\newblock URL
  \url{https://openaccess.thecvf.com/content_CVPR_2020/html/Su_Instance-Aware_Image_Colorization_CVPR_2020_paper.html}.

\bibitem[Wang et~al.(2020)Wang, Wen, Fu, Lin, Zou, Xue, and Zhang]{NeuralPose}
Jiashun Wang, Chao Wen, Yanwei Fu, Haitao Lin, Tianyun Zou, Xiangyang Xue, and
  Yinda Zhang.
\newblock Neural pose transfer by spatially adaptive instance normalization.
\newblock In \emph{IEEE/CVF Conference on Computer Vision and Pattern
  Recognition (CVPR)}, 2020.
\newblock URL
  \url{https://openaccess.thecvf.com/content_CVPR_2020/html/Wang_Neural_Pose_Transfer_by_Spatially_Adaptive_Instance_Normalization_CVPR_2020_paper.html}.

\bibitem[Wang et~al.(2018)Wang, Yu, Wu, Gu, Liu, Dong, Qiao, and
  Change~Loy]{Wang_2018_ECCV_Workshops}
Xintao Wang, Ke~Yu, Shixiang Wu, Jinjin Gu, Yihao Liu, Chao Dong, Yu~Qiao, and
  Chen Change~Loy.
\newblock {ESRGAN}: Enhanced super-resolution generative adversarial networks.
\newblock In \emph{European Conference on Computer Vision (ECCV) Workshops},
  2018.
\newblock URL
  \url{https://openaccess.thecvf.com/content_eccv_2018_workshops/w25/html/Wang_ESRGAN_Enhanced_Super-Resolution_Generative_Adversarial_Networks_ECCVW_2018_paper.html}.

\bibitem[Xudong et~al.(2017)Xudong, Qing, Haoran, Raymond Y.~K., and
  Zhen]{lsgan}
Mao Xudong, Li~Qing, Xie Haoran, Lau Raymond Y.~K., and Wang Zhen.
\newblock Least squares generative adversarial networks.
\newblock In \emph{IEEE/CVF International Conference on Computer Vision
  (ICCV)}, 2017.
\newblock URL
  \url{https://openaccess.thecvf.com/content_iccv_2017/html/Mao_Least_Squares_Generative_ICCV_2017_paper.html}.

\bibitem[Yang et~al.(2019)Yang, Hong, Jang, Zhao, and Lee]{diverseGan}
Dingdong Yang, Seunghoon Hong, Yunseok Jang, Tiangchen Zhao, and Honglak Lee.
\newblock Diversity-sensitive conditional generative adversarial networks.
\newblock In \emph{International Conference on Learning Representations
  (ICLR)}, 2019.
\newblock URL \url{https://arxiv.org/abs/1901.09024}.

\bibitem[Yang et~al.(2022)Yang, Jiang, Liu, and Loy]{yang2022unsupervised}
Shuai Yang, Liming Jiang, Ziwei Liu, and Chen~Change Loy.
\newblock Unsupervised image-to-image translation with generative prior.
\newblock In \emph{Conference on Computer Vision and Pattern Recognition},
  2022.

\bibitem[Zhang et~al.(2017)Zhang, Xu, Li, Zhang, Wang, Huang, and
  Metaxas]{zhang2017stackgan}
Han Zhang, Tao Xu, Hongsheng Li, Shaoting Zhang, Xiaogang Wang, Xiaolei Huang,
  and Dimitris~N Metaxas.
\newblock {StackGAN}: Text to photo-realistic image synthesis with stacked
  generative adversarial networks.
\newblock In \emph{{IEEE} International Conference on Computer Vision (ICCV)},
  2017.
\newblock URL
  \url{https://openaccess.thecvf.com/content_iccv_2017/html/Zhang_StackGAN_Text_to_ICCV_2017_paper.html}.

\bibitem[Zhang et~al.(2018)Zhang, Isola, Efros, Shechtman, and Wang]{LPIPS}
Richard Zhang, Phillip Isola, Alexei~A Efros, Eli Shechtman, and Oliver Wang.
\newblock The unreasonable effectiveness of deep features as a perceptual
  metric.
\newblock In \emph{IEEE Conference on Computer Vision and Pattern Recognition
  (CVPR)}, 2018.
\newblock URL
  \url{https://openaccess.thecvf.com/content_cvpr_2018/html/Zhang_The_Unreasonable_Effectiveness_CVPR_2018_paper.html}.

\bibitem[Zhang et~al.(2019)Zhang, Wang, Lin, and Qi]{srntt}
Zhifei Zhang, Zhaowen Wang, Zhe Lin, and Hairong Qi.
\newblock Image super-resolution by neural texture transfer.
\newblock In \emph{IEEE/CVF Conference on Computer Vision and Pattern
  Recognition (CVPR)}, 2019.
\newblock URL
  \url{https://openaccess.thecvf.com/content_CVPR_2019/html/Zhang_Image_Super-Resolution_by_Neural_Texture_Transfer_CVPR_2019_paper.html}.

\bibitem[Zhao et~al.(2017)Zhao, Mathieu, and LeCun]{EnegryGan}
Junbo~Jake Zhao, Micha{\"{e}}l Mathieu, and Yann LeCun.
\newblock Energy-based generative adversarial network.
\newblock In \emph{International Conference on Learning Representations
  (ICLR)}, 2017.
\newblock URL \url{http://arxiv.org/abs/1609.03126}.

\bibitem[Zheng et~al.(2018)Zheng, Ji, Wang, Liu, and Fang]{CrossCSR}
Haitian Zheng, Mengqi Ji, Haoqian Wang, Yebin Liu, and Lu~Fang.
\newblock {CrossNet}: An end-to-end reference-based super resolution network
  using cross-scale warping.
\newblock In \emph{European Conference on Computer Vision (ECCV)}, 2018.
\newblock URL
  \url{https://openaccess.thecvf.com/content_ECCV_2018/html/Haitian_Zheng_CrossNet_An_End-to-end_ECCV_2018_paper.html}.

\bibitem[Zhu et~al.(2017)Zhu, Park, Isola, and Efros]{cycleGan}
Jun-Yan Zhu, Taesung Park, Phillip Isola, and Alexei~A Efros.
\newblock Unpaired image-to-image translation using cycle-consistent
  adversarial networks.
\newblock In \emph{Conference on Computer Vision and Pattern Recognition
  (CVPR)}, 2017.
\newblock URL
  \url{https://openaccess.thecvf.com/content_iccv_2017/html/Zhu_Unpaired_Image-To-Image_Translation_ICCV_2017_paper.html}.

\end{thebibliography}

\section*{Supplementary Material}
%!TEX root = main.tex
\section{Qualitative Results}
\begin{figure*}
    \centering
    \includegraphics[width=0.68\textwidth]{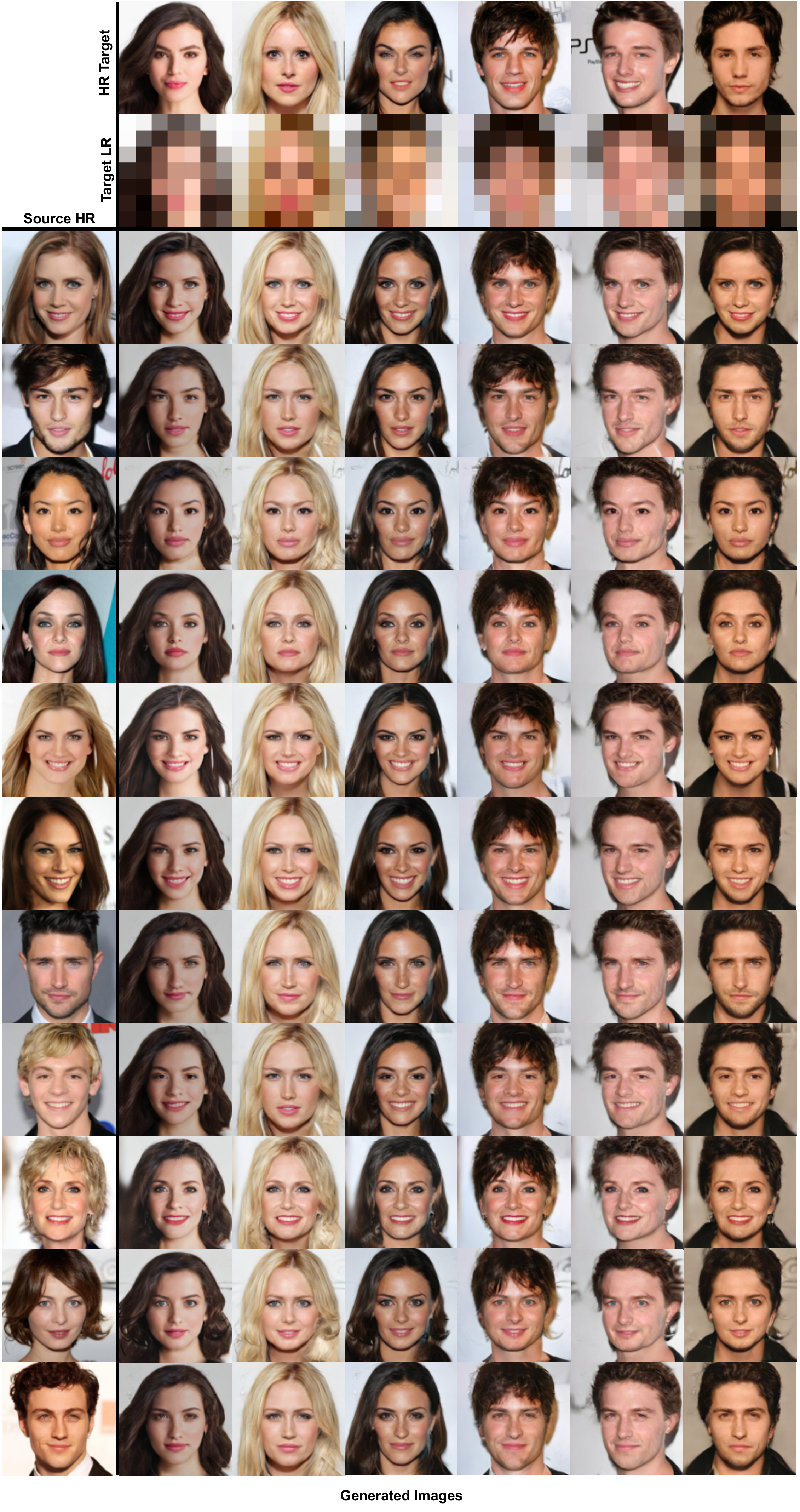}
    \caption{Generated samples on CelebA-HQ \citep{ProGanCelebA} conditioned on HR source (left column) and on the LR target (top row).}%
    \label{fig:add}%
\end{figure*}

\begin{figure*}
    \centering
    \includegraphics[width=0.94\textwidth]{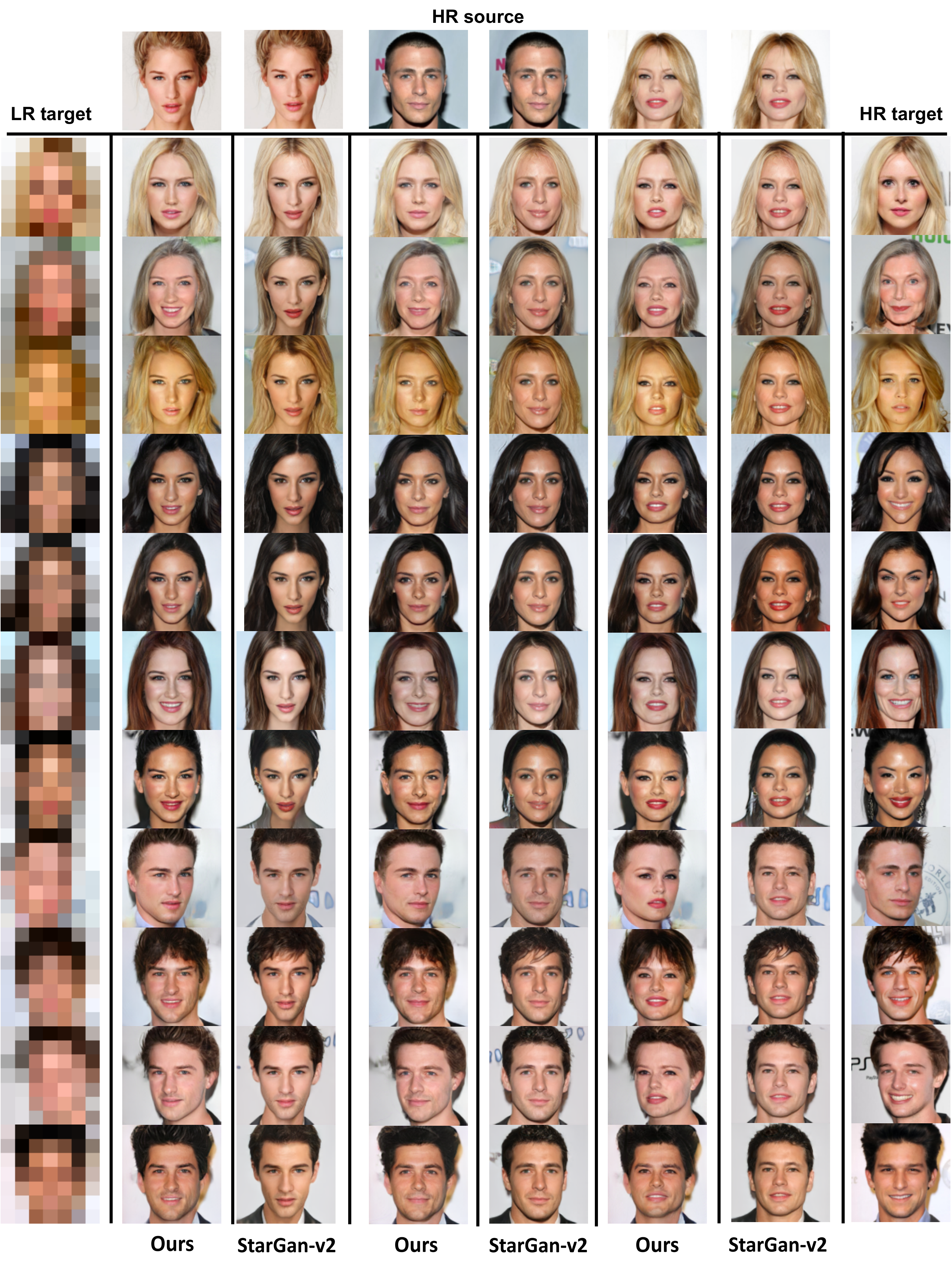}
    \caption{Comparison with StarGAN~v2 \citep{starganv} on CelebA-HQ \citep{ProGanCelebA}. To generate samples, our method uses LR target and HR source while StarGAN~v2 uses the HR source and HR target. 
    % \todo{Correct the way StarGAN~v2 is spelled/capitalized in the figure.}
    }%
    \label{fig:comp1}%
\end{figure*}

\begin{figure*}
    \centering
    \includegraphics[width=0.94\textwidth]{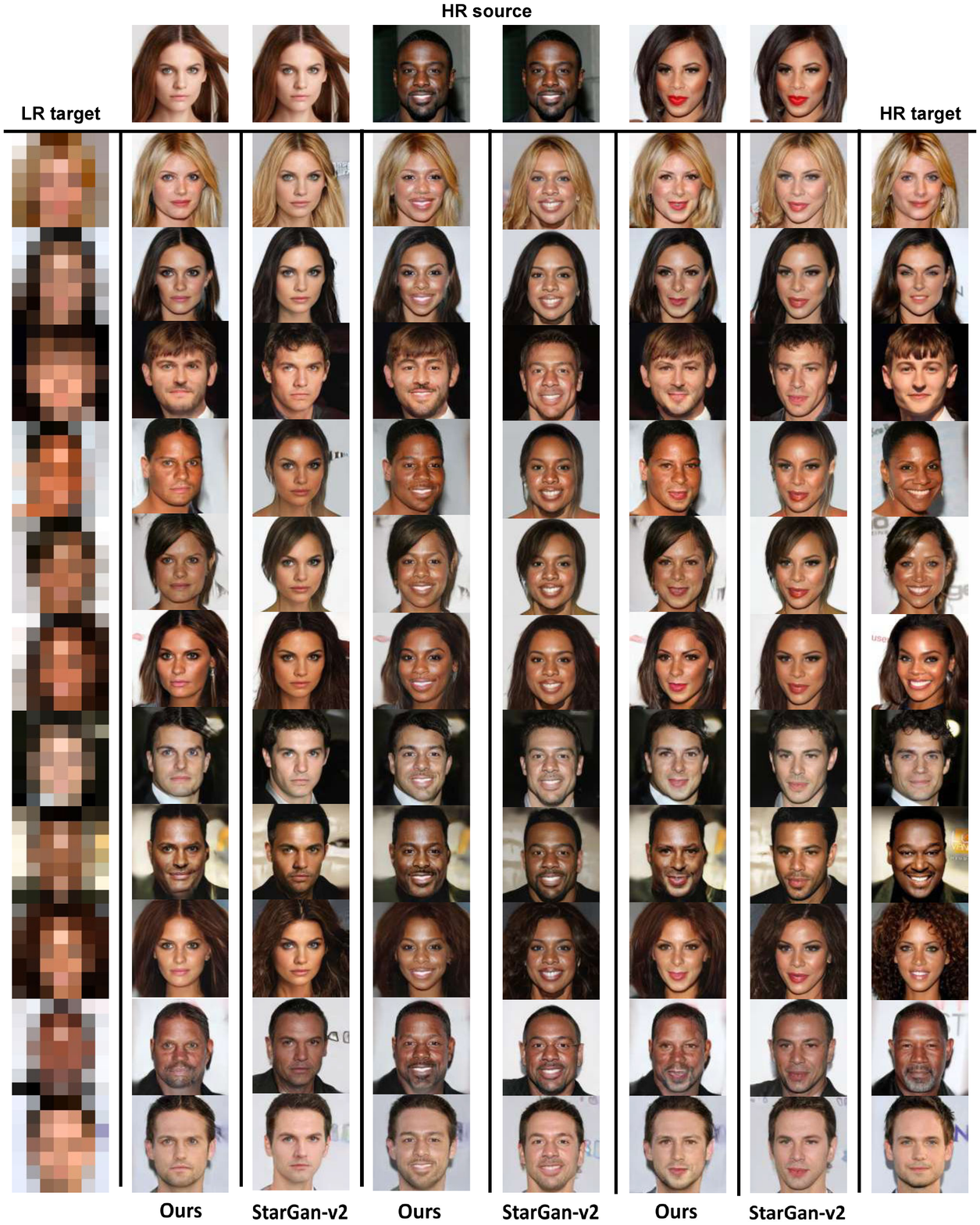}
    \caption{Comparison  with StarGAN~v2 \citep{starganv} on CelebA-HQ \citep{ProGanCelebA}. To generate samples, our method uses LR target and HR source while StarGAN~v2 uses the HR source and HR target. 
    % \todo{Correct the way StarGAN~v2 is spelled / capitalized in the figure.}
    }%
    \label{fig:comp2}%
\end{figure*}

\begin{figure*}
    \centering
    \includegraphics[width=0.9\textwidth]{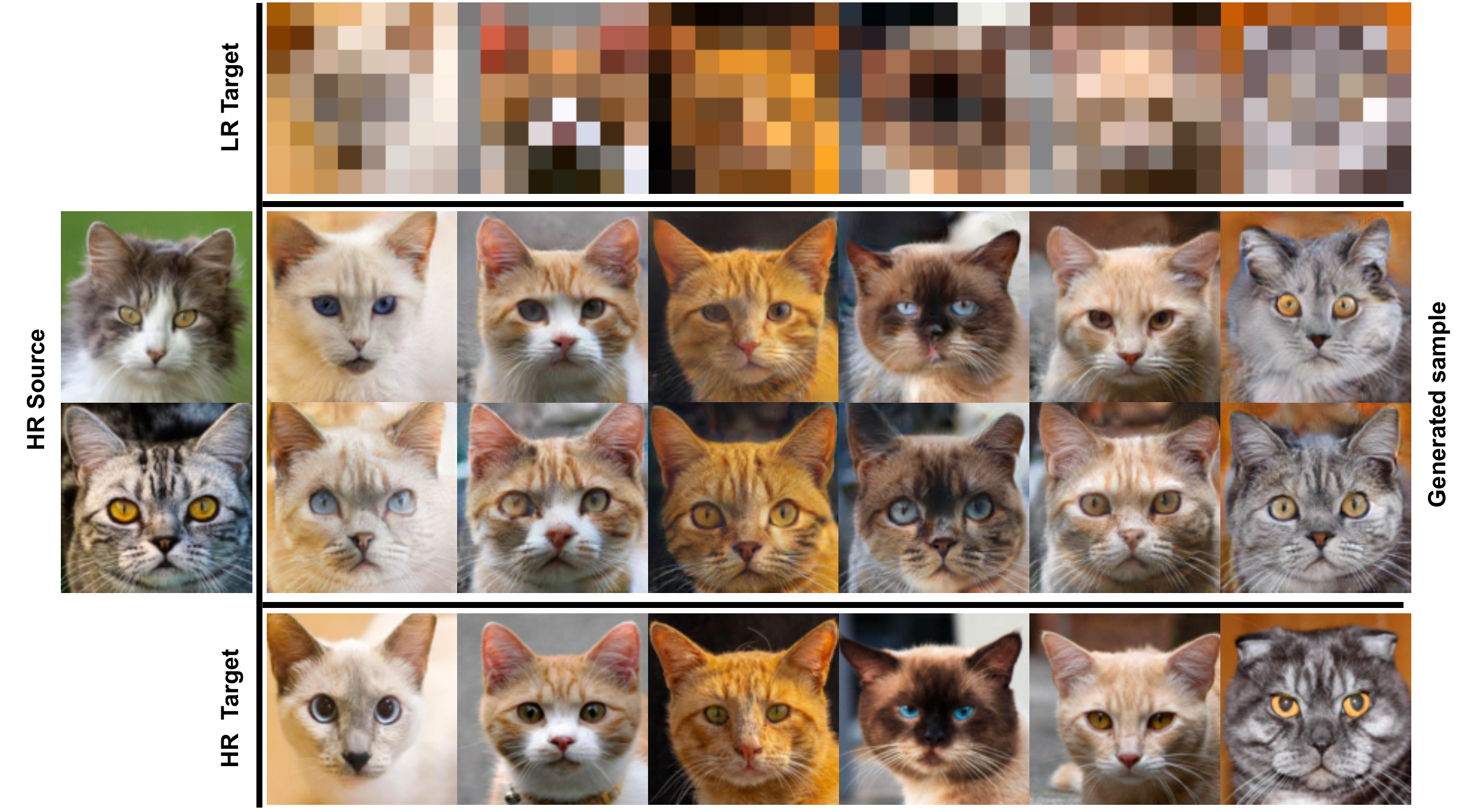}
    \caption{Generated samples on the domain cats of AFHQ \citep{stargan} conditioned on HR source (left column) and on the LR target (top row).}%
    \label{fig:cats}%
\end{figure*}

\begin{figure*}
    \centering
    \includegraphics[width=0.9\textwidth]{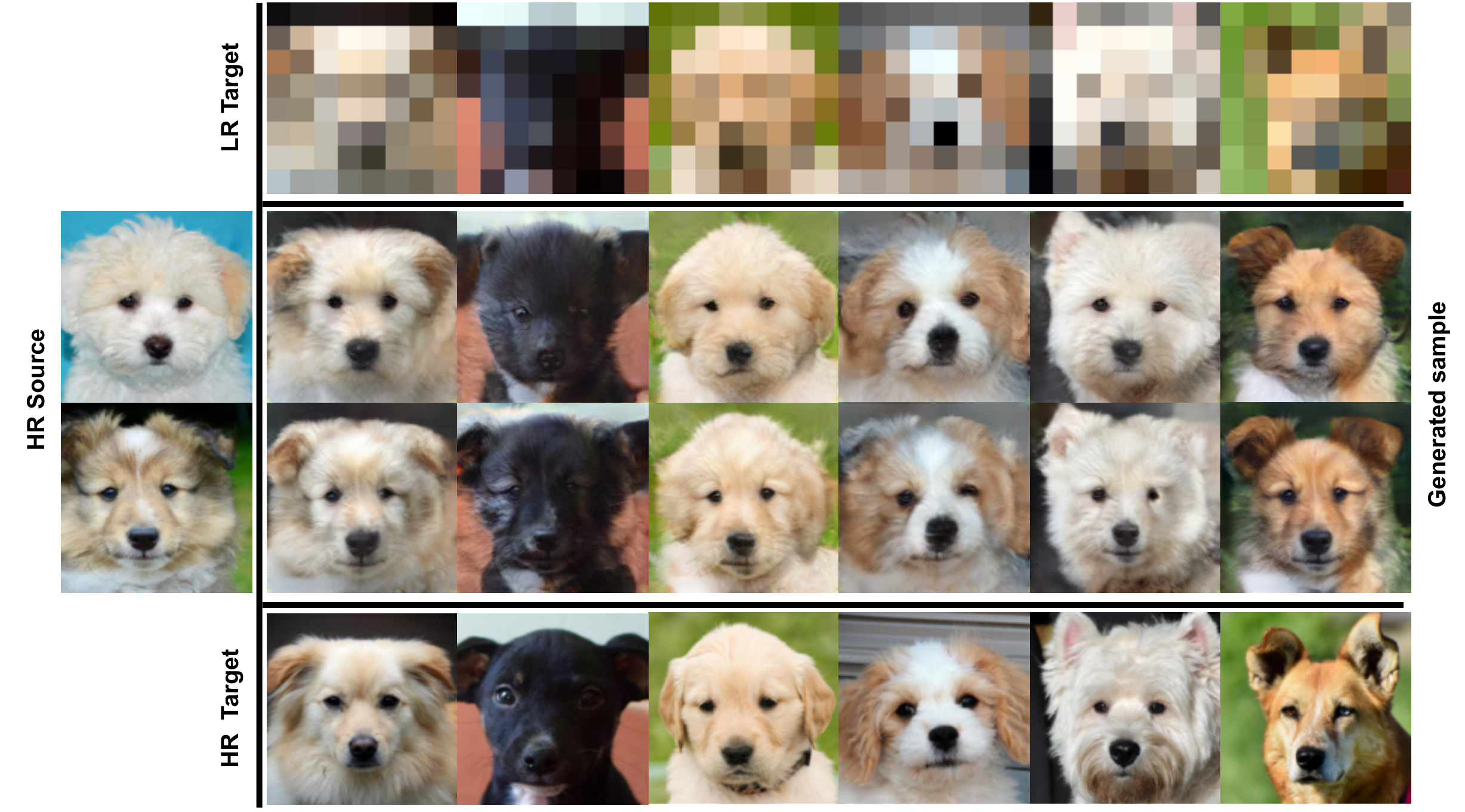}
    \caption{Generated samples on the domain dogs of AFHQ \citep{stargan} conditioned on HR source (left column) and on the LR target (top row).}%
    \label{fig:dogs}%
\end{figure*}

\begin{figure*}
    \centering
    \includegraphics[width=0.9\textwidth]{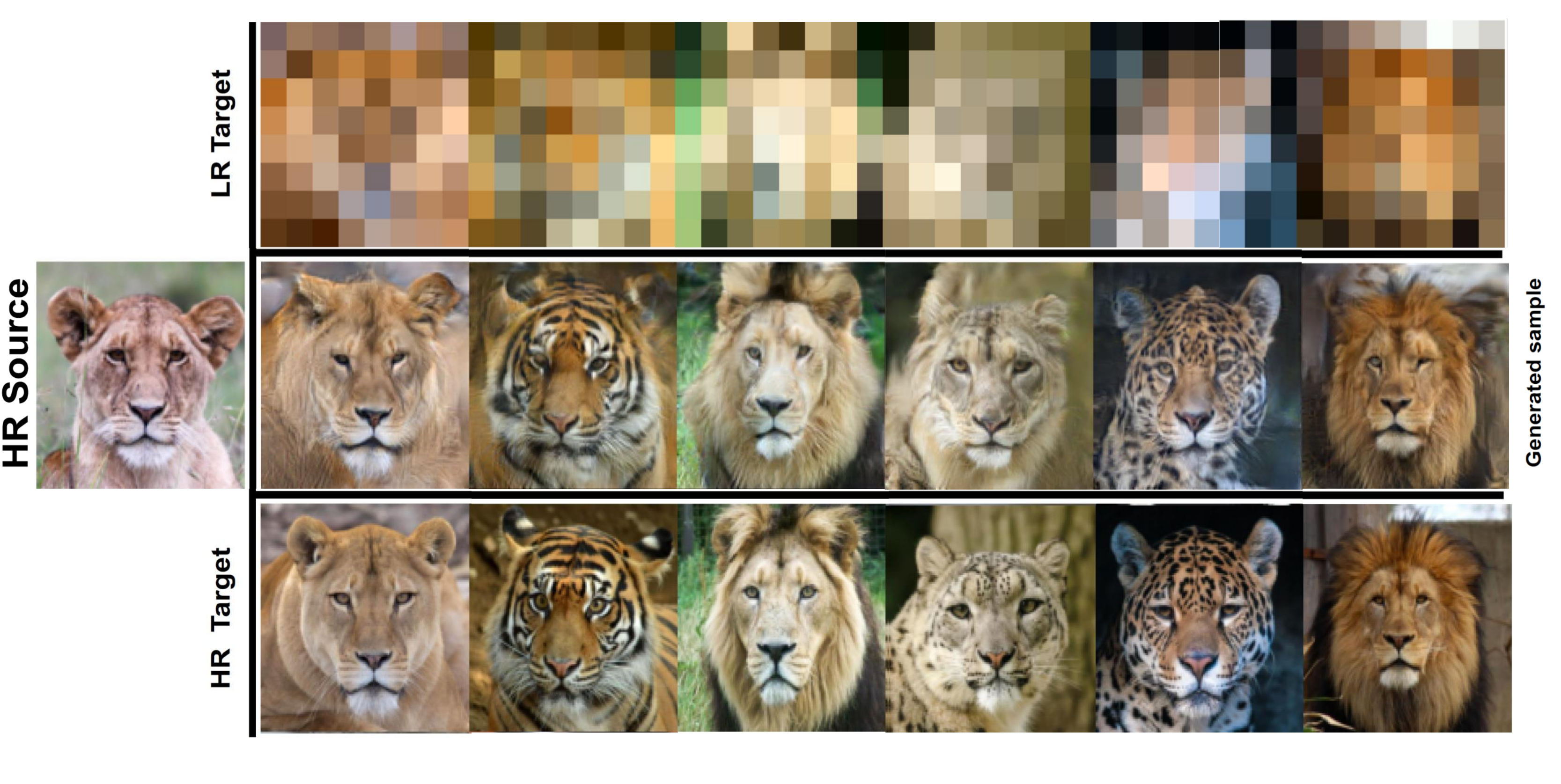}
    \caption{Generated samples on the domain wild of AFHQ \citep{stargan} conditioned on HR source (left column) and on the LR target (top row). 
    % \todo{There seems to be an error here, the images generated are the same over each column, there are duplicates over the two HR sources.}
    }%
    \label{fig:wild}%
\end{figure*}

We provide additional results on both CelebA-HQ and AFHQ (\cref{fig:add,fig:comp1,fig:comp2,fig:cats,fig:dogs,fig:wild}). On CelebA-HQ,  Our method generates diverse samples given the HR source and following the LR target. The network learns to follow the pose of the target LR image due to use of SPAdaIN \citep{NeuralPose}, while preserving the source features. We also provide additional comparison with reference guided StarGAN~v2 shown in \cref{fig:comp1,fig:comp2}. On AFHQ, we generate images on each domain separately, results are shown in \cref{fig:cats,fig:dogs,fig:wild}.

\section{Color Ablation}
\begin{figure*}
    \centering
    \includegraphics[width=0.8\textwidth]{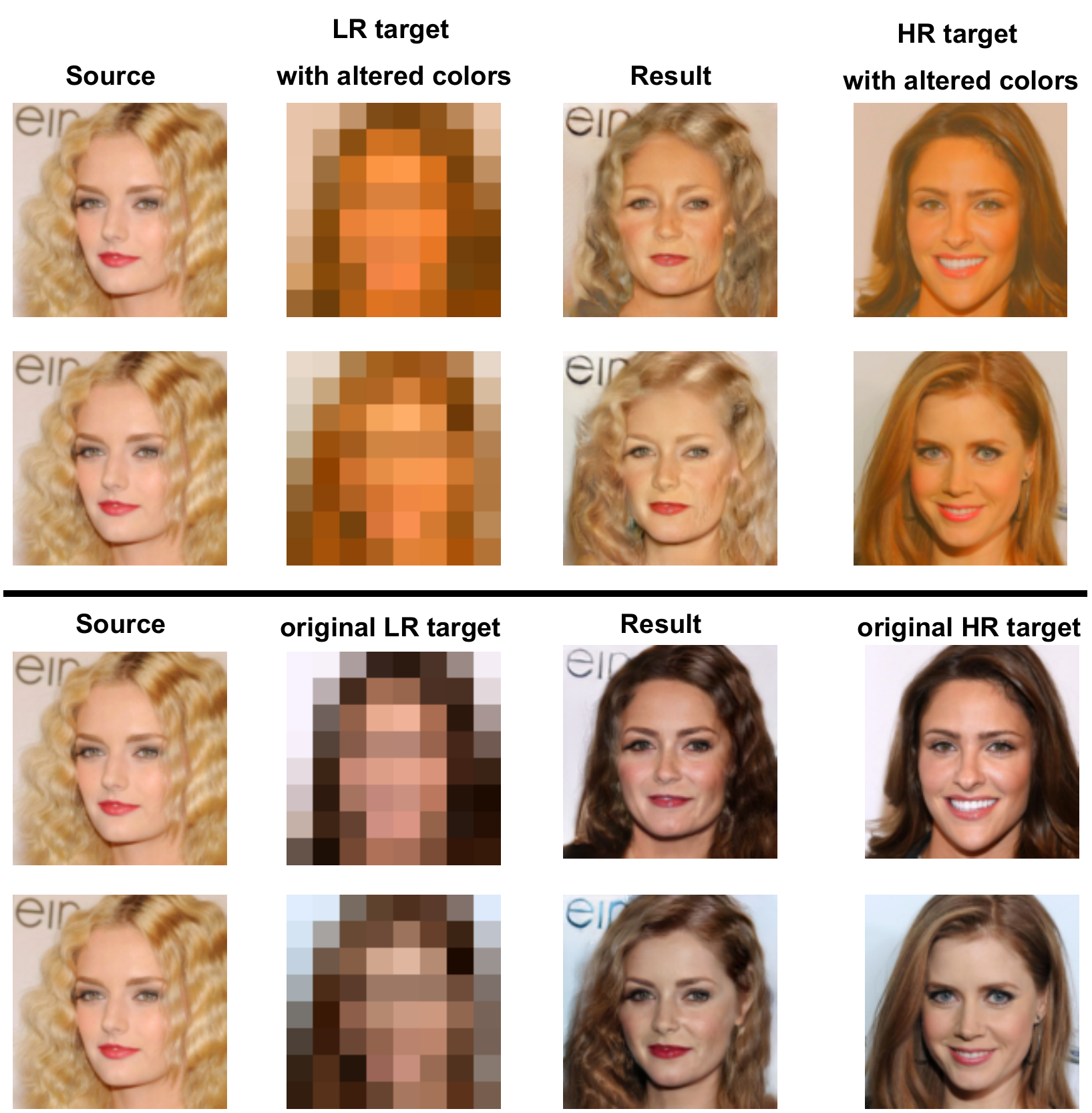}
    \caption{Generated samples on CelebA-HQ \citep{ProGanCelebA} while modifying the color of LR target using the source image.}%
    \label{fig:color1}%
\end{figure*}

To inspect the effect of color on the generation process, we choose to alter the original coloring of the target image, following color transfer method introduced in \cite{color}. In \cref{fig:color1}, we transfer the source image colors to the target image (HR and LR), and we use the target LR in the generation process. By comparing the generated image ``w/o'' and ``w/'' color transfer,  we notice that color does have impact on the generated images in terms of overall texture and evidently colors, a slight impact on the overall structure of the image.

\begin{figure*}
    \centering
    \includegraphics[width=0.8\textwidth]{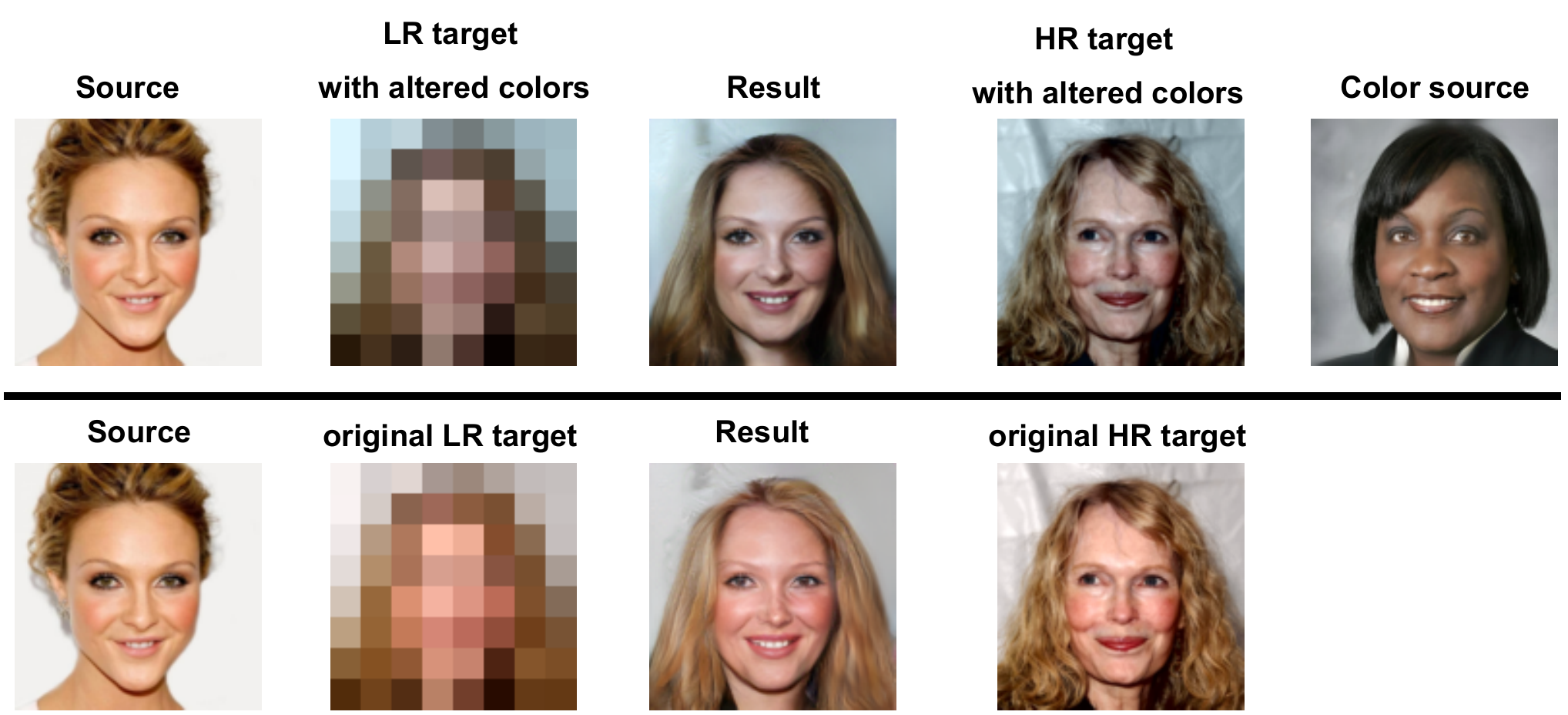}
    \caption{Generated samples on CelebA-HQ \citep{ProGanCelebA} while modifying the color of LR target using a third image.}%
    \label{fig:color2}%
\end{figure*}

In \cref{fig:color2}, we transfer the colors of a unrelated image, to the target image (HR and LR), and we use the target LR in the generation process. We observe that indeed the color does have a slight impact on the generation process.


\end{document}